\documentclass[aps,prd,twocolumn,superscriptaddress,amsmath,amssymb]{revtex4}
\pdfoutput=1
\usepackage{graphicx}
\usepackage{color}
\usepackage{ulem}
\usepackage{hyperref}
\usepackage{float}
\usepackage{cancel}
\usepackage{braket}
\usepackage{subfigure} 

\begin{document}

\title{Effects of the Violation of the Equivalence Principle at DUNE}
	\author{F.N. D\'iaz}
	\author{J. Hoefken}
	\author{A.M. Gago}
	\affiliation{Secci\'on F\'isica, Departamento de Ciencias, Pontificia Universidad Cat\'olica del Per\'u, Apartado 1761, Lima, Per\'u}	
	\begin{abstract}

A number of different effects of the violation of the Equivalence Principle (VEP), taken as sub-leading mechanism of neutrino flavor oscillation, are examined within the framework of the DUNE experiment. We study the possibility of obtaining a misleading neutrino oscillation parameter region caused by our unawareness of VEP. Additionally, we evaluate the impact on the measurement of CP violation and the distinction of neutrino mass hierarchy at DUNE. Besides, limits on VEP for a wide variety of textures of the matrix that connects neutrino gravity eigenstates to flavor eigenstates are imposed. An extra-task of our study is to set limits on Hamiltonian added terms considering different energy dependencies ($E^n$, with $n=0,1,2,3$) that can be associated to the usual Lorentz violating terms defined in the Standard Model Extension Hamiltonian. In order to understand our results, approximated analytical three neutrino oscillation probability formulae are derived.

	\end{abstract}
	\maketitle

\section{Introduction}	
The neutrino oscillation is caused by slight differences between neutrino masses (squared masses), which are already small in themselves, and the lack of coincidence between neutrino mass eigenstates and flavor eigenstates \cite{Fukuda:1998mi,Fukuda:2001nj,Ahmad:2002jz,Araki:2004mb,Adamson:2007gu,An:2012eh,Ahn:2012nd,Abe:2011fz,Kajita:2016vhj}. The long-distance interferometry characteristic of neutrino oscillations, in addition to their energy dependency, allows us to test sub-leading effects that can be produced by a variety of beyond standard oscillation physics such as non-standard interaction \cite{Gago:2001xg,Guzzo:2004ue,deGouvea:2015ndi,Masud:2016gcl,Liao:2016orc}, neutrino decay \cite{Frieman:1987as,Barger:1999bg,Bandyopadhyay:2002qg,Fogli:2004gy,Berryman:2014yoa,Picoreti:2015ika,Bustamante:2016ciw,Gago:2017zzy, Ascencio-Sosa:2018lbk,deSalas:2018kri}, quantum decoherence \cite{Lisi:2000zt,Barenboim:2006xt,Bakhti:2015dca,Carpio:2017nui,Capolupo:2018hrp,Carrasco:2018sca, Gomes:2020muc}, among others \cite{Adamson:2008aa,AguilarArevalo:2011yi,Li:2014rya}
. Nowadays, we are moving towards a neutrino oscillation physics precision era which implies that our sensitivity for performing searches for signatures from non-standard physics would be increased as well. One example of subleading non-standard physics that can be probed through oscillation physics is  the violation of Equivalence Principle (VEP). The Equivalence Principle is a central, heuristic principle that led Einstein to formulate his gravitation theory. In particular, the Weak Equivalence Principle states that, given a gravitational field, the trajectory followed by any falling body is independent of its mass. In the weak field limit, it says that in a given gravitational field all bodies fall in vacuum with the same acceleration, regardless of their masses. This is a manifestation of the equivalence between gravitational and inertial mass. The VEP mechanism, assuming massless neutrinos, was first introduced in order to explain the solar neutrino problem \cite{Gasperini:1988zf,Gasperini:1989rt,Halprin:1991gs,Pantaleone:1992ha,Butler:1993wi,Bahcall:1994zw,Mansour:1998nb,Gago:1999hi}; then, once the oscillation induced by mass was established as solution of the neutrino data, the studies involving VEP were reoriented in order to look for constraints on its parameters \cite{Yasuda:1994nu,Datta:2000hm,valdiviessotesis,Valdiviesso:2012nva,Esmaili:2014ota}. 

In this paper, we examine the potential of DUNE experiment \cite{Alion:2016uaj,Acciarri:2015uup} for imposing constraints on VEP parameters. Also we evaluate how its projected precision measurements of (sensitivity to) neutrino oscillation parameters could be affected by the presence of subleading VEP effects. In addition, we reinterpret our results beyond the context of VEP transforming its linear energy dependency into a quadratic, cubic, etc. In fact, we can make a correspondence between the aforementioned kind of terms with the Lorentz violating (LV) interaction terms appearing in the Standard Model Extension (SME) \cite{Colladay:1998fq,Colladay:1996iz}. The SME is a low-energy effective field theory that contains all possible LV operators, composed by ones originated from spontaneous Lorentz symmetry violation \cite{Kostelecky:1988zi}
 and others explicitly constructed.  

This paper goes as follows: in the second section we discuss the VEP theoretical framework. Then, in the third one, we make a full detailed description, at the level of probabilities, of the set of scenarios under study. In the fourth section, we present our findings. In the final section, we present our conclusions.   

\section{VEP Theoretical Framework}

The VEP is usuallly introduced through the breaking of the
universality of Newton's gravitational constant, $G_N$, being modified by a parameter $\gamma_i$ which depends on the mass of the $i$th-particle. As a result, a new constant $G'_{N} = \gamma_i G_N$ is defined, and, consequently, a mass-dependent gravity potential $\Phi' = \gamma_i \Phi$.

On the other hand, after replacing the space-time metric in the weak field approximation given by: $g_{\mu \nu}(x) = \eta_{\mu \nu} + h_{\mu \nu} (x)$, where $h_{\mu \nu} (x) = -2 \gamma_i \Phi(x)\delta_{\mu \nu}$ and $\eta_{\mu \nu}=diag(1,-1,-1,-1)$ is the Minkowski metric, in the relativistic invariant: $g_{\mu\nu} p^\mu p^\nu = m^2$, a modified energy-momentum relation is attained: $E^2 (1 -2 \gamma_i \Phi) = p^2 (1+2 \gamma_i \Phi) +m^2$ \cite{valdiviessotesis}. From the last relation, and taking $p^2  \gg m^2$ and neglecting terms $\Phi m^2/p$ and of $O(\Phi^2)$ we get:  
\begin{equation}
E_i \simeq p (1+2 \gamma_i \Phi) +\frac{m^2}{2p}  
\end{equation}
that leads us to the familiar expression:
\begin{equation}
\Delta E_{ij} = \frac{\Delta m^2}{2 E} + 2 E 
\Delta \gamma_{ij} 
\label{deltaEmn}
\end{equation}
where $\Delta \gamma_{ij}=\Phi (\gamma_i -\gamma_j)$. At the right hand side of the latter equation, the two contributions for the energy shift are shown: one due to the differences between neutrino mass eigenstates and the other one because of the differences between neutrino gravitational eigenstates. It is important to note that in the case of the mass-dependent VEP the neutrino gravitational eigenstates and the mass eigenstates are diagonal with respect to the same basis, being the general situation when these two types of eigenstates are assumed as not equal. Both aforementioned situations are treated in our analysis.

\subsection{Hamiltonian and oscillation probabilities}

The flavor basis Hamiltonian for three neutrino generation in matter is given by:
\begin{equation}
\textbf{H}_{\text{osc}}^{\text{f}} = \frac{1}{2E} \big[ \textbf{U} \textbf{H}_{\text{osc}}  \textbf{U}^\dagger +  \textbf{A}_{\text{matt}}\big] 
\end{equation}
with
\begin{eqnarray}
\textbf{H}_{\text{osc}}= \mathrm{diag}(0, \Delta m^2_{21}, \Delta m^2_{31}) \\
\textbf{A}_{\text{matt}}= \mathrm{diag}(A_{CC}, 0, 0)
\end{eqnarray}
where $A_{CC}=2\sqrt{2} G_{F} N_{e} E$. A generic Hamiltonian for the neutrino-gravitational eigenstates, written in the flavor basis, can be added to it:
\begin{equation}
\textbf{H}_{\text{osc}}^{\textbf{tot}} = \textbf{H}_{\text{osc}}^{\text{f}} +\textbf{H}_{\text g}^{\text f}
\label{htot} 
\end{equation}
with
\begin{eqnarray}
\textbf{H}_{\text g}^{\text f}= 2E \textbf{U}_{\text{g}} \textbf{H}_{\text{g}} \textbf{U}_{\text{g}}^\dagger\\
\textbf{H}_{\text g}= \mathrm{diag}(0, \Delta \gamma_{21}, \Delta \gamma_{31})
\end{eqnarray}
where $\textbf{U}$ is the usual PMNS matrix and $\textbf{U}_{\text g}$ is the analogous matrix that connects the neutrino-gravitational eigenstates to the flavor eigenstates. 
In order to get the matter oscillation probabilities formulae, that include perturbatively the gravitational effects, it is enough to take the formulae given in 
\cite{Liao:2016hsa}, developed in the context of Non-Standard Interactions, and make a careful replacement of the analogous terms. With this aim in hands, some definitions are presented to begin with. First,  $\textbf{V}_{\text g} = 2E \textbf{H}_{\text g}^{\text f} = 4E^{2}  \textbf{U}_{\text{g}} \textbf{H}_{\text{g}} \textbf{U}_{\text{g}}^\dagger$ where: 
\begin{equation}
\textbf{V}_{\text g} = k_E \left(\begin{matrix}
v_{ee} &v_{e\mu} e^{i \phi_{e\mu}} &v_{e\tau}e^{i \phi_{e\tau}} \\
v_{e\mu}e^{-i \phi_{e\mu}}&v_{\mu \mu}& v_{\mu \tau}e^{i \phi_{\mu \tau}} \\
v_{e\tau} e^{-i \phi_{e \tau}}&v_{\mu \tau}e^{-i \phi_{\mu \tau}}&v_{\tau \tau}
\end{matrix}\right)
\end{equation}
with $k_E = 4E^{2}$ (replace $k_E \equiv  A{'}$). We write  
$\textbf{U}_{\text{g}} \textbf{H}_{\text{g}} \textbf{U}_{\text{g}}^\dagger$  in terms of the generic matrix elements $v$, and their complex phases, with the purpose of having an easy match between these elements and their corresponding $\epsilon$ (and their phases) present in the prescription given in~\cite{Liao:2016hsa}. Then, we can rewrite Eq.(\ref{htot}):  
\begin{equation}
\textbf{H}_{\text{osc}}^{\textbf{tot}}  = \frac{1}{2E} \big[ 
\textbf{U} \textbf{H}_{\text{osc}}  \textbf{U}^\dagger +  \textbf{A}_{\text{matt}} +\textbf{V}_{\text g} \big] 
\end{equation}  
where:
\begin{equation}
\textbf{A}_{\text{matt}} +\textbf{V}_{\text g}=  k_E 
\left(\begin{matrix} \frac{A_{CC}}{k_E} +v_{ee} &v_{e\mu} e^{i \phi_{e\mu}} &v_{e\tau}e^{i \phi_{e\tau}} \\
v_{e\mu}e^{-i \phi_{e\mu}}&v_{\mu \mu}& v_{\mu \tau}e^{i \phi_{\mu \tau}} \\
v_{e\tau} e^{-i \phi_{e \tau}}&v_{\mu \tau}e^{-i \phi_{\mu \tau}}&v_{\tau \tau}
\end{matrix}\right)
\label{eq10}
\end{equation}
Thus, for getting the matter oscillation probability formulae it is necessary to replace $\frac{A_{CC}}{k_E} + v_{ee} \rightarrow 1+ \epsilon_{e e}  $ and $k_E \rightarrow A$, while for the rest $v \rightarrow \epsilon$ and $\phi \rightarrow \phi$ in Eq. (4) (Eq. (15)) given in \cite{Liao:2016hsa} (\cite{Majhi:2019tfi}) for the channels $\nu_\mu \rightarrow \nu_e$  ( $\nu_\mu \rightarrow \nu_\mu$). On top of these replacements we introduce the following notation: 
\begin{equation}
\begin{split}
&\tilde{A} = \frac{k_E}{EL} = \frac{4E}{L} \\
&\tilde{v}_{\alpha \beta} = E L v_{\alpha \beta} \\
&\tilde{A} \tilde{v}_{\alpha \beta} = k_E v_{\alpha \beta}
\end{split}
\end{equation}
where $L$ is the neutrino source-detector distance. Once all the aforementioned details are applied, the $\nu_\mu \rightarrow \nu_e$ oscillation probability turns out to be: 
\begin{equation}
\begin{split}
&P_{\nu_\mu \rightarrow \nu_e}^{\text{VEP}\bigoplus \text{SO}}\simeq P_{\nu_\mu \rightarrow \nu_e}^{\mathrm{SO}}  \\
& + 4\hat{A}\tilde{v}_{e\mu}   \left\lbrace xf \left[ s_{23}^{2}f\cos \left( \phi_{e\mu} + \delta\right) +c_{23}^{2}g\cos\left(\Delta+\delta+\phi_{e\mu}\right) \right] \right.\\ 
&\left. + yg \left[ c_{23}^{2} g \cos \phi_{e\mu} + s_{23}^{2} f \cos\left( \Delta-\phi_{e\mu}\right)\right]\right\rbrace \\ 
&+4\hat{A}\tilde{v}_{e\tau} s_{23}c_{23} \left\lbrace xf \left[ f\cos\left(\phi_{e\tau}+\delta\right) -g\cos\left(\Delta+\delta+\phi_{e\tau}\right) \right] \right.\\
&\left. -yg \left[g\cos\phi_{e\tau}-f\cos\left(\Delta-\phi_{e\tau}\right)\right] \right\rbrace\\
&+4\hat{A}^2g^2c_{23}^2|c_{23}\tilde{v}_{e\mu} e^{i \phi_{e\mu}}-s_{23}\tilde{v}_{e\tau} e^{i \phi_{e\tau}}|^2 \\ &+4\hat{A}^2f^2s_{23}^2|s_{23}\tilde{v}_{e\mu} e^{i \phi_{e\mu}}+c_{23}\tilde{v}_{e\tau} e^{i \phi_{e\tau}}|^2\\
&+8\hat{A}^2fgs_{23}c_{23} \left\lbrace c_{23}\cos \Delta \left[ s_{23} \left(\tilde{v}_{e\mu}^2-\tilde{v}_{e\tau}^2\right) \right.\right.\\
&\left.\left.+ 2c_{23}\tilde{v}_{e\mu}\tilde{v}_{e\tau}\cos\left(\phi_{e\mu}-\phi_{e\tau}\right)\right]-\tilde{v}_{e\mu}\tilde{v}_{e\tau}\cos\left(\Delta-\phi_{e\mu}+\phi_{e\tau}\right)\right\rbrace \\
&+ \mathcal{O}\left(s_{13}^2\tilde{v}_{\alpha \beta},s_{13}\tilde{v}_{\alpha \beta}^2,\tilde{v}_{\alpha \beta}^3\right) 
\end{split}
\label{numunue}
\end{equation} 
where
\begin{equation}
\begin{split}
&x=2s_{13}s_{23},\ \ y=2rs_{12}c_{12}c_{23}, \ \ r= |\Delta m_{21}^2/\Delta m_{31}^2|\\
&f,\bar{f}=\frac{\sin\left[\Delta\left(1\mp\hat{A}\tilde{v}'_{ee}\right)\right]}{1\mp\hat{A}\tilde{v}'_{ee}}, \ \ g=\frac{\sin \left(\hat{A}\tilde{v}'_{ee}\Delta\right)}{\hat{A}\tilde{v}'_{ee}} \\
&\tilde{v}'_{ee} =\frac{A}{\tilde{A}} + \tilde{v}_{ee},\ \ \Delta =\bigg|\frac{\Delta m_{31}^2 L}{4E}\bigg|, \ \ \hat{A}=\bigg|\frac{\tilde{A}}{\Delta m_{31}^2}\bigg| 
\end{split}
\label{defnumunue}
\end{equation}
and $s_{ij}=\sin \theta_{ij}$, $c_{ij}=\cos \theta_{ij}$. The antineutrino equation $\bar{\nu}_\mu \rightarrow \bar{\nu}_e$ is given by the Eq. (\ref{numunue}), changing $\hat{A}\rightarrow -\hat{A}$ (then $\bar{f}$ instead of $f$), $\delta \rightarrow -\delta$ and $\phi_{\alpha \beta} \rightarrow -\phi_{\alpha \beta}$. For the inverted hierarchy $\Delta \rightarrow - \Delta$, $y \rightarrow -y$ and $\hat{A} \rightarrow - \hat{A}$.
The $\tilde{v}_{\alpha \beta}$, one of the key parameters of  expansion, is $\sim \Delta \tilde{\gamma}_{ij} =E L \Delta \gamma_{ij}$. Our analytical probability formulae are valid as long as 
$\Delta \tilde{\gamma}_{ij}$ are taken to be not greater than $\mathcal{O}(0.1)$ in order to get less than 5$\%$ error between this analytical formula and the numerical one, within a neutrino energy ranging from 7 GeV to 14 GeV depending on the case. Other important parameters of expansion are the usual ones: $s_{13} \sim 0.1$ and $r \equiv |\Delta m_{21}^2/\Delta m_{31}^2| \sim 0.01$.

On the other hand, the oscillation probability for $\nu_\mu \rightarrow \nu_\mu$ disappearance channel is described by: 

{\small
	\begin{equation}
	\begin{split}
	&P_{\nu_\mu\rightarrow \nu_\mu}^{\text{VEP}\bigoplus \text{SO}} \simeq P_{\nu_\mu\rightarrow \nu_\mu}^{\mathrm{SO}} \\
	&-\tilde{v}_{\mu\tau}\hat{A}\cos\phi_{\mu\tau} \sin (2\theta_{23})\left[2\Delta s^2_{23} \sin(2\Delta)+4 \cos^2 (2\theta_{23}) \sin^2\Delta\right] \\
	&+\hat{A}\left(\tilde{v}_{\mu \mu}-\tilde{v}_{\tau \tau}\sin^2 (2\theta_{23}) \cos (2\theta_{23})\right)\left[\Delta \sin(2\Delta)-2\sin^2\Delta\right] \\
	&+\mathcal{O}(r,s_{13},\tilde{v}_{\alpha \beta}^2)
	\end{split}
	\label{numunumu}
	\end{equation}}

It is important to note that we have rewritten the probabilities in such a way that the pure standard oscillation contribution, $P_{\nu_\alpha\rightarrow \nu_\beta}^{\mathrm{SO}}$,  is separated from those terms which mixed the new physics parameters and the standard ones. Additionally, whenever we use these analytical oscillation probabilities formulae, the $P_{\nu_\alpha\rightarrow \nu_\beta}^{\mathrm{SO}}$ term is numerically calculated. This is done in order to achieve a better agreement between these (semi) analytical probabilities and those fully numerically calculated.

\begin{table}[!h]\begin{center}\begin{tabular}{c|c|c}
			Parameter & Value & Error \\\hline
			$\theta_{12}$ & $33.62^{\circ}$ & $0.77^{\circ}$\\
			$\theta_{13} (\mathrm{NH})$ & $8.54^{\circ}$ & $0.15^{\circ}$\\
			$\theta_{23} (\mathrm{NH})$ & $47.2^{\circ}$ & $1.9^{\circ}$ \\
			$\Delta m_{21}^{2} $ & $7.4 \times 10^{-5} \mathrm{eV}^{2}$ & $0.2\times 10^{-5} \mathrm{eV}^{2}$ \\
			$\Delta m_{31}^{2}(\mathrm{NH}) $ & $2.494 \times 10^{-3} \mathrm{eV}^{2}$ & $0.032\times 10^{-3} \mathrm{eV}^{2}$ \\
			Baseline & $1300 \mathrm{Km}$ & -\\
		\end{tabular}
		\caption{DUNE baseline and values for standard oscillation parameters taken from \cite{Nufit} (January 2018).}
		\label{ParametersOscillation}
\end{center}\end{table}	

\subsection{Lorentz violation interpretation}
\label{LV}
Before we proceed it is worthwhile to mention that
	the VEP prescription presented here, and its posterior results, can be reinterpreted for a general energy exponent case. The latter can be implemented since the only parameter that encodes the VEP effects in our probability formulation is $\Delta \tilde{\gamma}_{ij} =E L \Delta \gamma_{ij}$. Therefore, it is enough to replace: $ 2E \rightarrow E^n \implies E  \rightarrow E^n/2$ where $n$ can be any number, which is equivalent to replace $ \bf H^f_g \propto 2E \rightarrow \bf H^f_g \propto E^n $, in order to make our probability formulae able to test a power-law energy dependency,  for a given exponent,  and, accordingly,  with the chance of reinterpreting the results that we present here for a general situation. The cases when $n=0,1,2,..$ match with the isotropic Lorentz violating terms described in the effective Hamiltonian of the SME \cite{Aartsen:2017ibm}, the minus sign in some coefficients can be reabsorbed in $\Delta \gamma_{ij}$.

\section {Violation of Equivalence Principle scenarios}
In this section, we study a set of VEP cases corresponding to different choices for $\textbf{U}_{\text{g}}$
and $\Delta \tilde{\gamma}_{ij} (=E L \Delta \gamma_{ij})$, deriving their specific oscillation 
probabilities from our general formulae given in Eq.(\ref{numunue})
and Eq.(\ref{numunumu}). For a direct and simple understanding of a given case, these specific formulae should be a much shorter version of the general one. Our simplification criteria is to preserve only the most relevant terms responsible for the main patterns of behavior of a given case.
 
\subsection{$\textbf{U}_{\text g}=\textbf{U}$}
The simplest case to study is when we take the PMNS matrix $\textbf{U}$ equal to $\textbf{U}_{\text g}$. Considering the mixing angles and the $\Delta \tilde{\gamma}_{ij}$, the $\tilde{v}_{\alpha\beta}$ and $\phi_{\alpha\beta}$ are explicitly written for $\nu_\mu \rightarrow \nu_e$ and $\nu_\mu \rightarrow \nu_\mu$ keeping the coefficients of order 
not greater than $s_{13}^2 \Delta \tilde{\gamma}_{ij}$ or $r \Delta \tilde{\gamma}_{ij}$ or  $s_{13} \Delta \tilde{\gamma}_{ij}^2$, i.e. only up to $\mathcal{O}(0.001)$, given that $s_{13}\sim \mathcal{O}(0.1)$, $r \sim \mathcal{O}(0.01)$ and $\Delta \tilde{\gamma}_{ij} \sim \mathcal{O}(0.1)$. For the neutrino appearance channel $\nu_\mu \rightarrow \nu_e$, 

\begin{equation}
\tilde{v}_{ee}= c_{13}^2s_{12}^2\Delta \tilde{\gamma}_{21}+s_{13}^2\Delta \tilde{\gamma}_{31}
\label{vee}
\end{equation}

\begin{equation}
\begin{split}
\phi_{e\mu}= \arctan\left[\frac{\sin\delta(k_2\Delta \tilde{\gamma}_{21}-k_3\Delta \tilde{\gamma}_{31})}{(k_1-k_2\cos \delta)\Delta \tilde{\gamma}_{21} +k_3 \cos \delta \Delta \tilde{\gamma}_{31}}\right]
\end{split}
\end{equation}

\begin{equation}
\tilde{v}_{e\mu}=\frac{(k_1-k_2\cos \delta)\Delta \tilde{\gamma}_{21} +k_3 \cos \delta \Delta \tilde{\gamma}_{31}}{\cos\phi_{e\mu}}
\end{equation}

\begin{equation}
\phi_{e\tau}=\arctan\left[\frac{-\sin\delta(k_1'\Delta \tilde{\gamma}_{21}+k_3'\Delta \tilde{\gamma}_{31})}{(k_2'+k_1'\cos \delta)\Delta \tilde{\gamma}_{21}+k_3' \cos \delta \Delta \tilde{\gamma}_{31}}\right]
\end{equation}

\begin{equation}
\tilde{v}_{e\tau}=\frac{(k_2'+k_1'\cos \delta)\Delta \tilde{\gamma}_{21}+k_3' \cos \delta \Delta \tilde{\gamma}_{31}}{-\cos \phi_{e\tau}}
\end{equation}

where

\begin{equation}
\begin{split}
&k_1=c_{12}c_{23}c_{13}s_{12},\ k_2=s_{12}^2s_{13}s_{23}c_{13}, \ k_3=c_{13}s_{23}s_{13}\\
&k_1'=s_{12}^2c_{23}s_{13}c_{13}, \ k_2'=c_{12}c_{13}s_{12}s_{23},\ k_3'= c_{13}c_{23}s_{13}
\end{split}
\end{equation}

For the neutrino disappearance channel $\nu_\mu \rightarrow \nu_\mu$

\begin{equation}
\begin{split}
\tilde{v}_{\mu \mu}= \Delta \tilde{\gamma}_{31} c_{13}^2 s_{23}^2 + \Delta \tilde{\gamma}_{21} [c_{12}^2c_{23}^2+s_{12}^2s_{13}^2s_{23}^2 \\
 -2c_{12}c_{23}s_{12}s_{13}s_{23}\cos\delta]
\end{split}
\label{vuu}
\end{equation}

\begin{equation}
\begin{split}
\tilde{v}_{\tau \tau}= \Delta \tilde{\gamma}_{31} c_{13}^2 c_{23}^2 + \Delta \tilde{\gamma}_{21} [c_{12}^2s_{23}^2+s_{12}^2s_{13}^2c_{23}^2 \\
-2c_{12}s_{23}s_{12}s_{13}c_{23}\cos\delta]
\end{split}
\label{vtt}
\end{equation}

\begin{equation}
\phi_{\mu \tau}= \arctan\left[\frac{\sin \delta (f_3 + f_4) \Delta \tilde{\gamma}_{21}}{\Delta \tilde{\gamma}_{31} f_1 + \Delta \tilde{\gamma}_{21} [f_2 + \cos\delta(f_3-f_4)]} \right]
\label{phiut}
\end{equation}

\begin{equation}
\tilde{v}_{\mu \tau}=\frac{\Delta \tilde{\gamma}_{31} f_1 + \Delta \tilde{\gamma}_{21} [f_2 + \cos\delta(f_3-f_4)]}{\cos \phi_{\mu \tau}}
\label{vut}
\end{equation}

with

\begin{equation}
\begin{array}{r@{}l}
&f_1=c_{13}^2s_{23}c_{23}, \ f_2=c_{23}s_{12}^2s_{13}^2s_{23}-c_{12}^2s_{23}c_{23} \\
&f_3=c_{12}s_{23}^2s_{12}s_{13}, \ f_4 = c_{12} c_{23}^2 s_{12} s_{13}
\end{array}
\end{equation}

In the following calculations, and within the scenario $\textbf{U}_{\text g}=\textbf{U}$, two cases are studied: ($\Delta \gamma_{21} = 0 \neq \Delta \gamma_{31} $) and ($\Delta \gamma_{21}\neq 0 =\Delta \gamma_{31}$).

\subsubsection{Case 1}

In this case, $\Delta \gamma_{21} = 0$ and $\Delta \gamma_{31} \neq 0$, the expression  for $\nu_\mu \rightarrow \nu_e$ is:
\begin{equation}
\label{eqs1c1me}
\begin{split}
P_{\nu_\mu \rightarrow \nu_e}^{\text{VEP}\bigoplus \text{SO}} \simeq & \ P_{\nu_\mu \rightarrow \nu_e}^{\text{SO}} + C_{1} s_{13}^2 \Delta \tilde{\gamma}_{31}
\end{split}
\end{equation}
		
\begin{equation}
\begin{split}
C_{1} = & 8 f^2 s_{23}^2 / \Delta \\
\end{split}
\end{equation}
meanwhile, the $\nu_\mu \rightarrow \nu_\mu$ disappearance channel is given by: 
\begin{equation}
\label{nunus1c1me}
\begin{split}
P_{\nu_\mu \rightarrow \nu_\mu}^{\text{VEP}\bigoplus \text{SO}} \simeq & \ P_{\nu_\mu \rightarrow \nu_\mu}^{\text{SO}} - \sin 2\Delta \sin^2 2\theta_{23} \Delta \tilde{\gamma}_{31}
\end{split}
\end{equation}			
			
\begin{figure}[h!]
\includegraphics[scale=0.55]{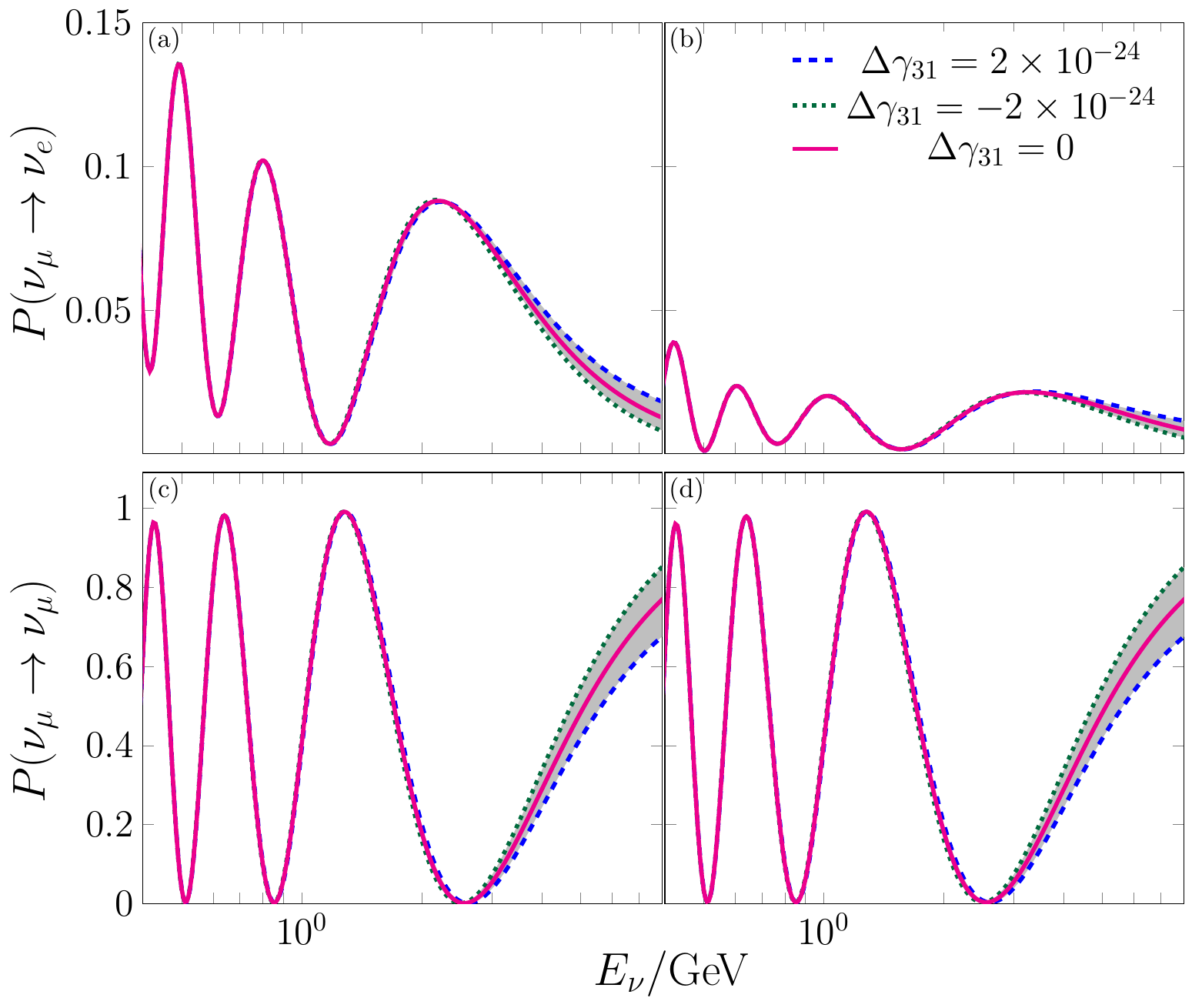}
\caption{Oscillation probability depending on the neutrino energy and considering scenario A/case 1. Figures (b) and (d) represent the $\bar{\nu}_e$ appearance and $\bar{\nu}_\mu$ disappearance  oscillation probability, respectively. We consider $\delta = -\pi/2$ and $L=1300 \ \mathrm{km}$.}	
\label{probs1c1mm}
\end{figure}

In Fig.~\ref{probs1c1mm} we can see that there are slight differences between $\text{VEP}\bigoplus \text{SO}$ and pure SO in the $\nu_\mu \rightarrow \nu_e$ appearance channel along the energy range. In turn, the impact is a bit more significant in the $\nu_\mu \rightarrow \nu_\mu$ disappearance channel. The higher differences in the $\nu_\mu \rightarrow \nu_\mu$ disappearance channel  can be explained by the presence of terms of orders $\Delta \tilde{\gamma}_{31} \sim 
\mathcal{O}(0.1)$ in  Eq.~(\ref{nunus1c1me}). While, the minor discrepancies in $\nu_\mu \rightarrow \nu_e$ are because
only terms scaled by $s_{13}^2 \Delta \tilde{\gamma}_{31}\sim  \mathcal{O}(0.001)$ are appearing in Eq.~(\ref{eqs1c1me}). This contribution has the same sign of $\Delta \tilde{\gamma}_{31}$, 
regardless it is a neutrino or an antineutrino due to the absence of $\delta_{\text{CP}}$ in that term. In the case of the channel $\nu_\mu \rightarrow \nu_\mu$ the contribution is negative respect to the sign of $\Delta \tilde{\gamma}_{31}$ and it is independent of being neutrino or antineutrino (there is no
$\delta_{\text{CP}}$ in the corresponding term).

\subsubsection{Case 2}

In this case, $\Delta \gamma_{21} \neq 0$ and $\Delta \gamma_{31} = 0$, the expression for $\nu_\mu \rightarrow \nu_e$ is:

\begin{equation}
\label{eqs1c2me}
\begin{split}
P_{\nu_\mu \rightarrow \nu_e}^{\text{VEP}\bigoplus \text{SO}} \simeq & \ P_{\nu_\mu \rightarrow \nu_e}^{\text{SO}} + C_{1} \cos \delta_{\mathrm{CP}} s_{13} \Delta \tilde{\gamma}_{21} \\ &- C_{2} \sin \delta_{\mathrm{CP}} s_{13} \Delta \tilde{\gamma}_{21} \\ &+ C_{3} r \Delta \tilde{\gamma}_{21} - C_{4} s_{13}^2 \Delta \tilde{\gamma}_{21} + C_{5} (\Delta \tilde{\gamma}_{21})^2
\end{split}
\end{equation}
		
with:
	
\begin{equation}
\begin{split}
C_{1} = & 8 f g \cos \Delta s_{12} c_{12} s_{23} c_{23} / \Delta \\
C_{2} = & 8 f g \sin \Delta s_{12} c_{12} s_{23} c_{23} / \Delta \\
C_{3} = & 8 g^2 s^2_{12} c^2_{12} c^2_{23} / \Delta \\
C_{4} = & 8 f^2 s^2_{12} s^2_{23} / \Delta \\
C_{5} = & 4 g^2 s^2_{12} c^2_{12} c^2_{23} / \Delta^2
\end{split}
\end{equation}
														
where the survival probability of $\nu_{\mu} \rightarrow \nu_{\mu}$ is:
\begin{equation}
\begin{split}
P_{\nu_\mu \rightarrow \nu_\mu}^{\text{VEP}\bigoplus \text{SO}} \simeq & \ P_{\nu_\mu \rightarrow \nu_\mu}^{\text{SO}} + \sin 2\Delta c^2_{12} \sin^2 2\theta_{23} \Delta \tilde{\gamma}_{21}
\end{split}
\end{equation}
\begin{figure}[h!]
\includegraphics[scale=0.55]{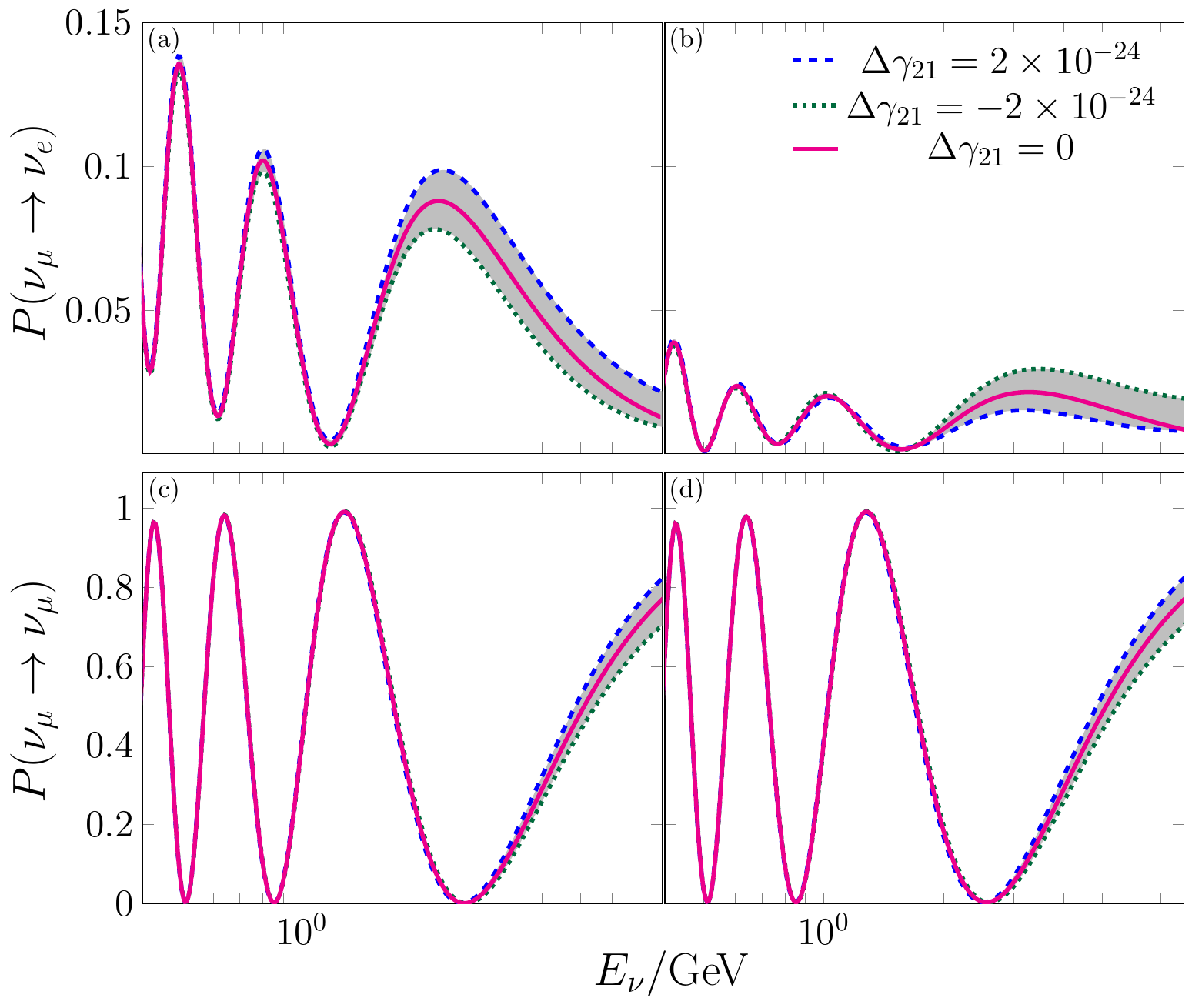}
\caption{Oscillation probability depending on the neutrino energy and considering scenario A/case 2. Figures (b) and (d) represent the $\bar{\nu}_e$ appearance and $\bar{\nu}_\mu$ disappearance  oscillation probability, respectively. We consider $\delta = -\pi/2$ and $L=1300 \ \mathrm{km}$.}
\label{probsAc2mm}	
\end{figure}
In contrast to the former case, and as it is shown in Fig.~\ref{probsAc2mm}, higher differences between $\text{VEP}\bigoplus \text{SO}$ and SO are registered for the $\nu_\mu \rightarrow \nu_e$ channel than for the case of  the $\nu_\mu \rightarrow \nu_\mu$ channel. In the $\nu_\mu \rightarrow \nu_e$ channel, the increment of the discrepancy, respect to the former case, relies on the fact that in this probability there are  terms of order of $s_{13} \Delta \tilde{\gamma}_{21}\sim  \mathcal{O}(0.01)$. The sign of the overall contribution is positive (negative) for neutrinos and $\Delta \gamma_{21}>0$ (antineutrinos and $\Delta \gamma_{21}<0$). The neutrino/antineutrino sign dependency occurs because of the emergence of $\delta_{\text{CP}}$ in the dominant terms of the contribution (note that the term associated to $C_1$ vanishes given that  
$\delta_{\text{CP}}=-\pi/2$). For the $\nu_\mu \rightarrow \nu_\mu$ channel, despite there is a term scaled for $\Delta \tilde{\gamma}_{21} \sim 
\mathcal{O}(0.1)$, the unlikeness is less noticeable, in comparison to the transition channel, since the contribution of this term is just smaller, by contrast with the magnitude of $\ P_{\nu_\mu \rightarrow \nu_\mu}^{\text{SO}}$,  than the corresponding 
ones for the transition channel.       

On the other hand, it is interesting to note, that the probabilities for the degenerate case, $\Delta \gamma_{21} =\Delta \gamma= \Delta \gamma_{31}$, can be attained simply by replacing $s^2_{12} \rightarrow c^2_{12}$ in $C_{4}$. The behavior of the relative differences between probabilities are rather similar than those shown here for the general case.  

\subsection{$\textbf{U}_{\text g} \neq \textbf{U}$}

Under the condition $\textbf{U}_{\text g} \neq \textbf{U}$, we develop three cases, which are selected according to 
three different choices of texture for the mixing matrix of the gravity eigenstates, $\textbf{U}_{\text g}$. Each texture is denoted by $\textbf{U}^{ij}_{\text g}$ which means that ${\theta^g_{ij}}$ is 
the only angle set as different from zero in this matrix.    
\subsubsection{Texture $\theta_{13}$}

\begin{figure}[h!]
\includegraphics[scale=0.55]{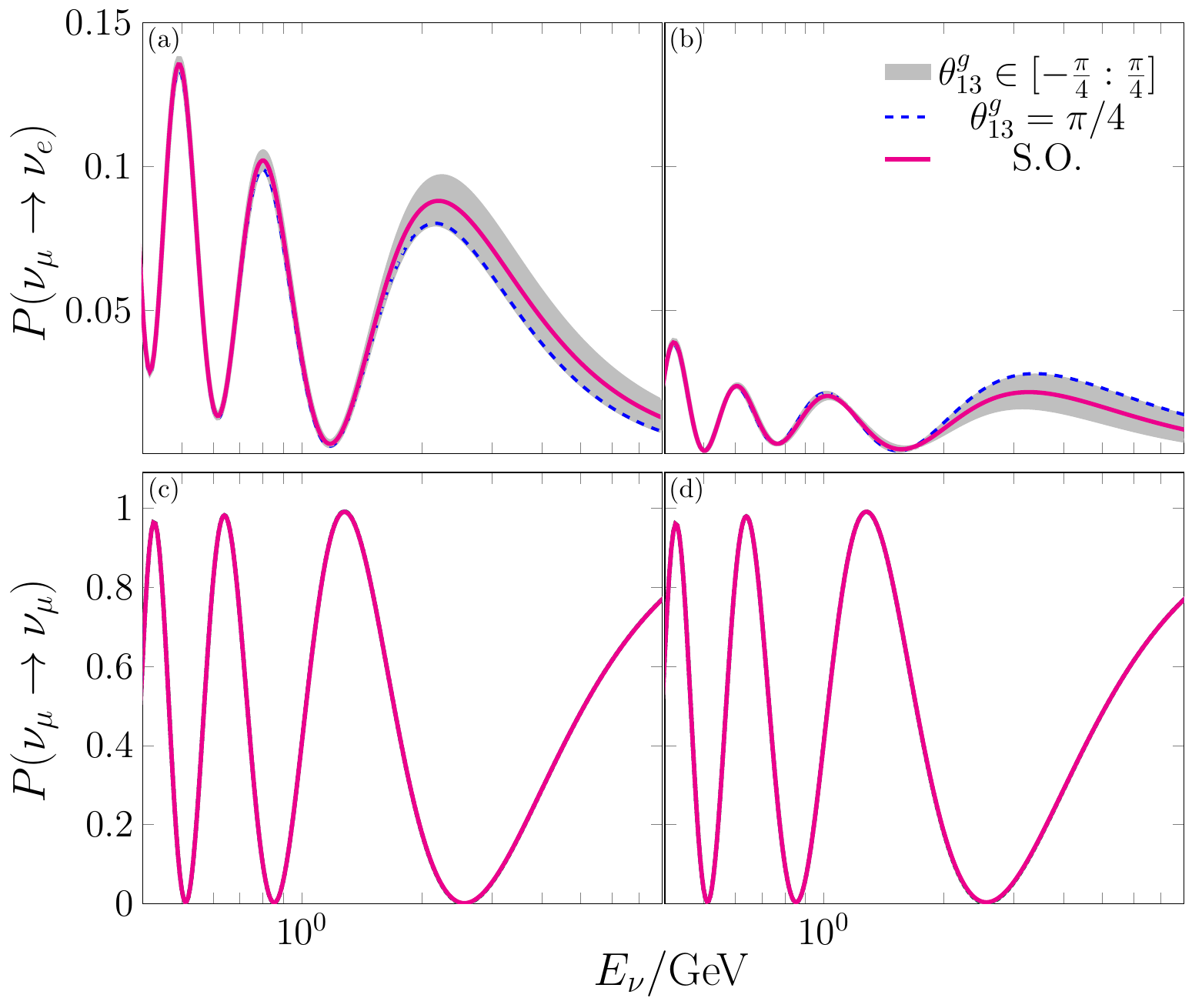}
\caption{Oscillation probability depending on the neutrino energy and considering scenario B/texture $\theta_{13}$. Figures (b) and (d) represent the $\bar{\nu}_e$ appearance and $\bar{\nu}_\mu$ disappearance oscillation probability, respectively. We consider $\Delta \gamma_{31}= 2 \times 10^{-24}$, $\delta = -\pi/2$ and $L=1300 \ \mathrm{km}$.}	
\label{nue_sc2_case1}
\end{figure}

The $\textbf{U}_{\text g}$ matrix for this case is given by
\begin{equation}
\textbf{U}^{13}_{\text g}=\left(\begin{matrix}
c^g_{13} & 0 & s^g_{13}\\
0 & 1 & 0 \\
- s^g_{13} & 0 & c^g_{13}
\end{matrix}\right)
\end{equation}
where $c^g_{ij} \equiv \cos \theta_{ij}^g$ and $s^g_{ij} \equiv \sin \theta_{ij}^g$. To select $\theta^g_{13} \neq 0$ implies a two generation reduction of the probability formula keeping only  $\Delta \gamma_{31}$, from the gravitational sector. After the proper replacements and simplifications the $\nu_\mu \rightarrow \nu_e$ oscillation channel 
takes the following form:
\begin{equation}
\label{eqs2c1me}
\begin{split}
P_{\nu_\mu \rightarrow \nu_e}^{\text{VEP}\bigoplus \text{SO}}
 \simeq & \ P_{\nu_\mu \rightarrow \nu_e}^{\text{SO}} \\& + C_{1} \cos \delta_{\mathrm{CP}} s_{13} \Delta \tilde{\gamma}_{31} + C_{2} \sin \delta_{\mathrm{CP}} s_{13} \Delta \tilde{\gamma}_{31} \\ &- C_{3} r \Delta \tilde{\gamma}_{31} + C_{4} (\Delta \tilde{\gamma}_{31})^2
\end{split}
\end{equation}
where:
\begin{equation}
\begin{split}
C_{1} = & 8 f (f - g \cos \Delta) s_{23}^2 c_{23} s^g_{13} c^g_{13} \ \Delta \\
C_{2} = & 8 fg \sin \Delta s_{23}^2 c_{23} s^g_{13} c^g_{13} / \Delta \\
C_{3} = & 8 g (g - f \cos \Delta) s_{12} c_{12} s_{23} c_{23}^2 s^g_{13} c^g_{13} / \Delta \\
C_{4} = &4 (f^2 + g^2 - 2 f g \cos \Delta) s^2_{23} c^2_{23} s^{g \ 2}_{13} c^{g \ 2}_{13} / \Delta^2
\end{split}
\label{eqs2c1meC}
\end{equation}
In the same way, the $\nu_\mu \rightarrow \nu_\mu$ disappearance channel is given by:
\begin{equation}
\label{eqs2c1mm}
\begin{split}
P_{\nu_\mu \rightarrow \nu_\mu}^{\text{VEP}\bigoplus \text{SO}}
 \simeq & \ P_{\nu_\mu \rightarrow \nu_\mu}^{\text{SO}} - \frac{2}{\Delta} \sin \Delta (\Delta \cos \Delta - \sin \Delta) \\ & \times s_{23} c_{23} \sin 4\theta_{23} c^{g \ 2}_{13} \Delta \tilde{\gamma}_{31}
\end{split}
\end{equation}
As it is observed in Fig.~\ref{nue_sc2_case1} 
the differences in the $\nu_\mu \rightarrow \nu_e$ channel
are of the same order than in the last case, 
which is because of the appearance in the  
probability of terms  $s_{13} \Delta \tilde{\gamma}_{31}\sim  \mathcal{O}(0.01)$, similar to those in Eq.~(\ref{eqs1c2me}). Since here   
 $\Delta \gamma_{31}$ is taken as positive, the sign 
of the overall contribution depends only on them being neutrinos (negative) or antineutrinos (positive). Also, as it can be extrapolated from the probability, the maximum disparity with respect to 
the SO is arising when $\theta_{13}^g = \pm \pi/4$, because it maximizes/minimizes $\sin 2\theta_{13}^g$. The divergences 
between the $\nu_\mu \rightarrow \nu_\mu$ probabilities are 
negligible because of  
the term containing VEP is proportional to 
$\sin 4\theta_{23} \sim 0$, recalling that $\theta_{23}$ is close to $\pi/4$.  

\subsubsection{Texture $\theta_{12}$}
For this texture the $\textbf{U}_{\text g}$ is given by:
\begin{equation}
\textbf{U}^{12}_{\text g}=\left(\begin{matrix}
c^g_{12} & s^g_{12} & 0 \\
-s^g_{12} & c^g_{12} & 0 \\
0&0&1
\end{matrix}\right)
\end{equation}
Here the expression for the $\nu_\mu \rightarrow \nu_e$ appearance channel turns out to be:
\begin{equation}
\label{eqs2c2me}
\begin{split}
P_{\nu_\mu \rightarrow \nu_e}^{\text{VEP}\bigoplus \text{SO}} \simeq & \ P_{\nu_\mu \rightarrow \nu_e}^{\text{SO}} \\& + C_{1} \cos \delta_\mathrm{CP} s_{13} \Delta \tilde{\gamma}_{21} - C_{2} \sin \delta_\mathrm{CP} s_{13} \Delta \tilde{\gamma}_{21} \\ &+ C_{3} r \Delta \tilde{\gamma}_{21} + C_{4} ( \Delta \tilde{\gamma}_{21})^2
\end{split}
\end{equation}

where:

\begin{equation}
\begin{split}
C_{1} = & 8 f (f s^2_{23} + g c^2_{23} \cos \Delta) s_{23} s^g_{12} c^g_{12} / \Delta \\
C_{2} = & 8 f g \sin \Delta s_{23} c^2_{23} s^g_{12} c^g_{12} / \Delta \\
C_{3} = & 8  g (f s^2_{23} \cos \Delta + g c^2_{23}) s_{12} c_{12} c_{23} s^g_{12} c^g_{12} / \Delta \\
C_{4} = & 4  (f^2 s^4_{23} + g^2 c^4_{23} + 2 f g s^2_{23} c^2_{23} \cos \Delta) s^{g \ 2}_{12} c^{g \ 2}_{12}  / \Delta^2
\end{split}
\label{eqs2c2meC}
\end{equation}
On the other hand, the $\nu_\mu \rightarrow \nu_\mu$ disappearance channel is:
\begin{equation}
\label{eqs2c2mm}
\begin{split}
P_{\nu_\mu \rightarrow \nu_\mu}^{\text{VEP}\bigoplus \text{SO}} \simeq & \ P_{\nu_\mu \rightarrow \nu_\mu}^{\text{SO}} + \frac{2}{\Delta} \sin \Delta (\Delta \cos \Delta-\sin \Delta) \\ &\times s_{23} c_{23} \sin 4\theta_{23} c^{g \ 2}_{12} \Delta \tilde{\gamma}_{21}
\end{split}
\end{equation}

\begin{figure}[h!]
\includegraphics[scale=0.55]{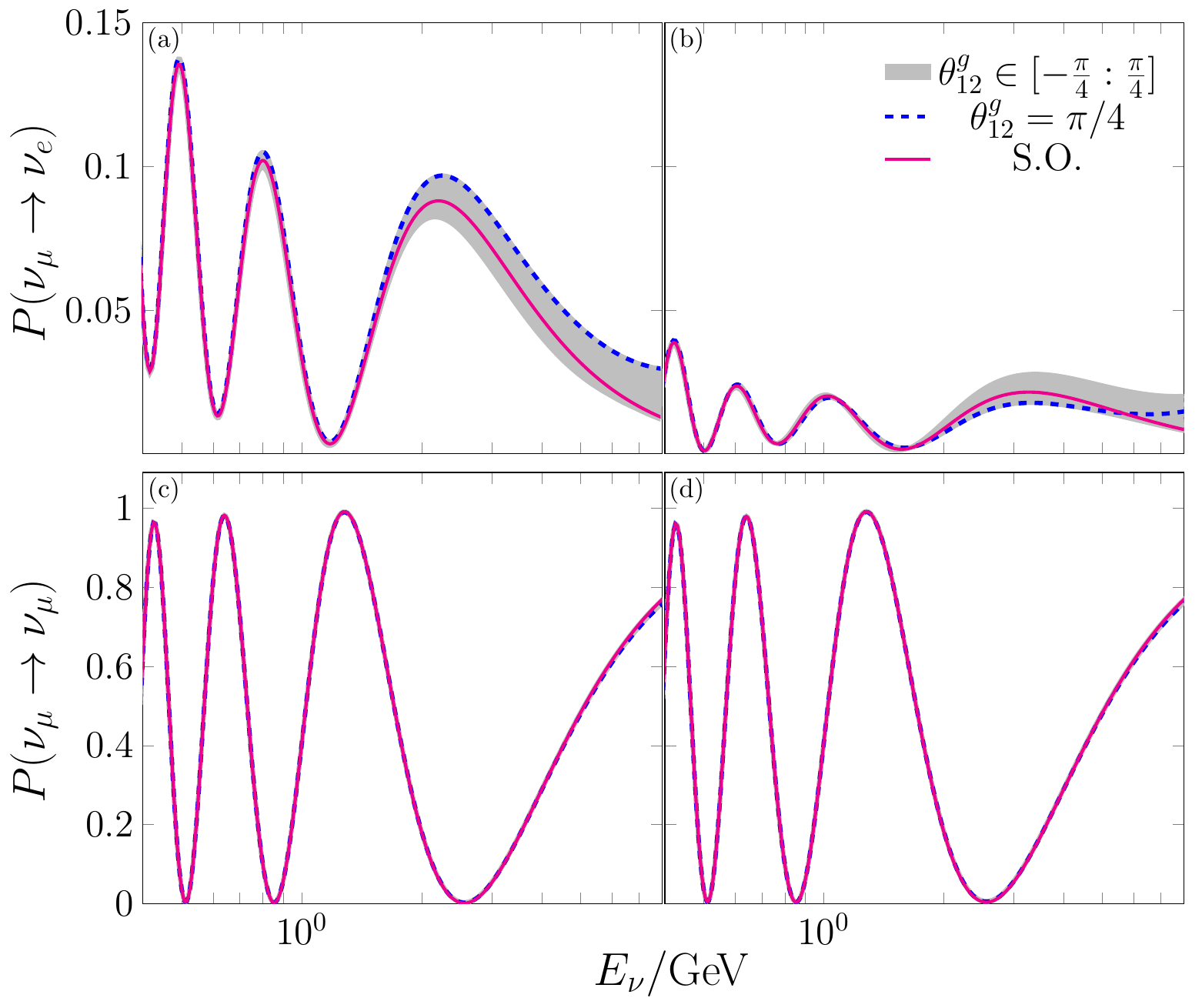}
\caption{Oscillation probability depending on the neutrino energy and considering scenario B/texture $\theta_{12}$. Figures (b) and (d) represent the $\bar{\nu}_e$ appearance and $\bar{\nu}_\mu$ disappearance oscillation probability, respectively. We consider $\Delta \gamma_{21}= 2 \times 10^{-24}$, $\delta = -\pi/2$ and $L=1300 \ \mathrm{km}$.}	
\label{nue_sc2_case2}
\end{figure}

As it can be seen in Fig.~\ref{nue_sc2_case2}, the pattern of the probabilities are akin to those presented in the former case, which    
is reasonable to expect in light of the similarities in the formulae for both cases. Therefore, 
parallel arguments used for explaining the previous case can be applied here. The only change is that the sign of the overall contribution,  that distinguish $\text{VEP}\bigoplus \text{SO}$ from $\text{SO}$, is positive for neutrinos and negative for antineutrinos in the $\nu_\mu \rightarrow \nu_e$ channel for this case. In the channel $\nu_\mu \rightarrow \nu_\mu$, as before, the differences between $\text{VEP}\bigoplus \text{SO}$ and $\text{SO}$ are negligible.     

\subsubsection{Texture $\theta_{23}$}
Here, our selection for the texture of $\textbf{U}_{\text g}$ goes as follows:
\begin{equation}
\textbf{U}^{23}_{\text g}=\left(\begin{matrix}
1&0&0 \\
0& c^g_{23} & s^g_{23} \\
0&-s^g_{23}&c^g_{23}
\end{matrix}\right)
\end{equation}
Since the $\Delta\gamma_{23}$ can be written as a function of  $\Delta\gamma_{31}$ and $\Delta\gamma_{21}$, we subdivide, this particular texture, into two different sub-cases.   

\paragraph{$\Delta\gamma_{21} = 0$ and $\Delta\gamma_{31} \neq 0$}\hfill 
\label{scenarioBsubcase3a}

It can be checked from Eq. (\ref{numunue}) that, for the  $\nu_\mu \rightarrow \nu_e$ channel all the perturbative contributions up to $\mathcal{O}(10^{-3})$ vanish. Meanwhile, the $\nu_\mu \rightarrow \nu_\mu$ has non-null perturbative contribution at $\Delta \tilde{\gamma}_{31} \sim \mathcal{O}(0.1)$, where its expression turns out to be as follows:
\begin{equation}
\begin{split}
P_{\nu_\mu \rightarrow \nu_\mu}^{\text{VEP}\bigoplus \text{SO}} \simeq & \ P_{\nu_\mu \rightarrow \nu_\mu}^{\text{SO}} - \frac{4}{\Delta} \Big(\Delta \cos \Delta \sin 2\theta_{23} \cos (2(\theta_{23}-\theta^g_{23})) \\ &- \sin \Delta \cos 2\theta_{23} \sin (2(\theta_{23}-\theta^g_{23})) \Big) \\ &\times \sin \Delta s_{23} c_{23} \Delta \tilde{\gamma}_{31}
\end{split}
\end{equation}

\paragraph{$\Delta\gamma_{21} \neq 0$ and $\Delta\gamma_{31} = 0$}\hfill
\label{scenarioBsubcase3b}

As the case above, for the $\nu_\mu \rightarrow \nu_e$ appearance channel there is no pertubative contribution 
up to terms scaled by factors of $\mathcal{O}(10^{-3})$, which represents 
an almost zero contribution. Likewise, the $\nu_{\mu} \rightarrow \nu_{\mu}$ channel has non-negligible perturbative contribution:
\begin{equation}
\label{eqs2c3bmm}
\begin{split}
P_{\nu_\mu \rightarrow \nu_\mu}^{\text{VEP}\bigoplus \text{SO}} \simeq & \ P_{\nu_\mu \rightarrow \nu_\mu}^{\text{SO}} + \frac{2}{\Delta} \Big( \sin \Delta (\Delta \cos \Delta - \sin \Delta) \\ &\times \sin 4\theta_{23} \cos 2\theta_{23}^g + \big(2 \sin^2 \Delta \cos^2 2\theta_{23} \\ &+ \Delta \sin 2\Delta \sin^2 2\theta_{23} \big) \sin 2\theta_{23}^g \Big) s_{23} c_{23} \Delta \tilde{\gamma}_{21}
\end{split}
\end{equation}
\begin{figure}[h!]
\includegraphics[scale=0.55]{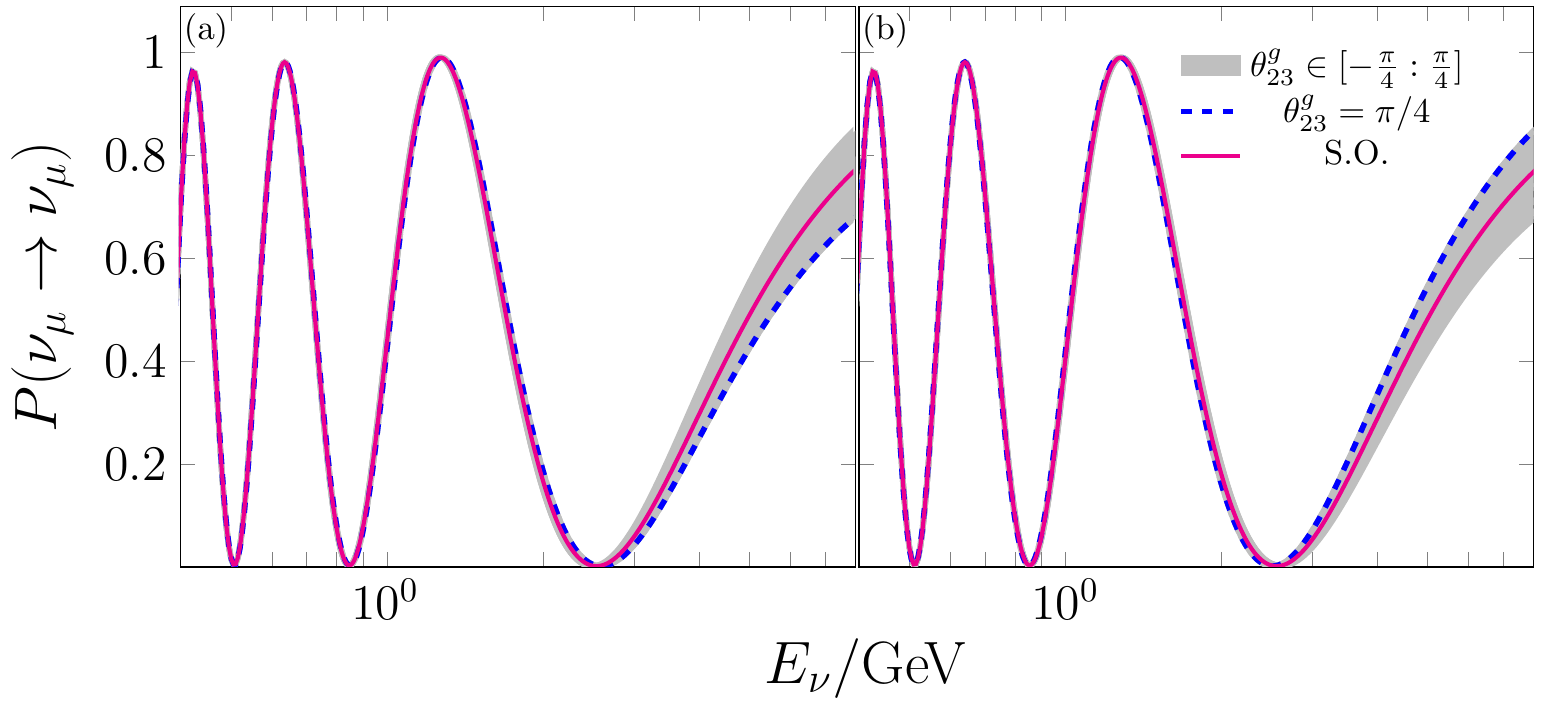}
\caption{Oscillation probability depending on the neutrino energy and considering scenario B/texture $\theta_{23}$. Figures (a) and (b) represent the sub-cases  \textit{a} and \textit{b}, respectively. We consider $\Delta \gamma_{21}= 2 \times 10^{-24}$ for sub-case \textit{a}, $\Delta \gamma_{31}= 2 \times 10^{-24}$ for sub-case \textit{b}, $\delta = -\pi/2$ and $L=1300 \ \mathrm{Km}$.}	
\label{numu_sc2_case3}
\end{figure}

In Fig.~\ref{numu_sc2_case3}, where it is only plotted the 
$\nu_{\mu} \rightarrow \nu_{\mu}$ channel, it is possible to note 
appreciable discrepancies of similar magnitudes for 
the sub-cases \textit{a} and \textit{b} between $\text{VEP}\bigoplus \text{SO}$
and $\text{SO}$. The magnitudes of these discrepancies 
are similar for 
both sub-cases but opposite in sign. For sub-case \textit{a}, 
the VEP contribution is negative while for \textit{b} it is positive. Additionally, for both sub-cases, as in the 
textures $\theta^g_{13}$ and $\theta^g_{12}$, it is confirmed  
that the utmost divergence (maximization of the VEP effect) is 
reached when $\theta^g=\pm \pi/4$. Furthermore, the 
probabilities for neutrinos are only displayed 
in Fig.~\ref{numu_sc2_case3} since their counterpart for antineutrinos are identical.


%
\section{Simulation and Results}

In the simulations the inputs from \cite{Alion:2016uaj} are used considering the optimized fluxes and an exposure of $3.5$ years for neutrino and antineutrino mode, Forward Horn current (FHC) and Reverse Horn Current (RHC) respectively. The default configuration of signal and background given by the DUNE collaboration (\cite{Alion:2016uaj} and \cite{Acciarri:2015uup}) is also used.

Throughout the present work, the values in Table \ref{ParametersOscillation} are considered as the current best fit values (CBFV). Given that the probability distributions are non-Gaussian, especially for $\theta_{23}$, the uncertainty  is calculated dividing by 6 the $3\sigma$ allowed region for each parameter. Because the $\delta_{\mathrm{CP}}$ is not sufficiently constrained, no priors are used, though an importance to $-\pi/2$ is considered because it is the closest value to the best fit \cite{Nufit}.

The GLoBES package is used to simulate DUNE \cite{Huber:2004ka,Huber:2007ji}. In this context, the following definition of $\chi^{2}$ \cite{Carpio:2018gum} is regarded:

\begin{equation}
\chi^{2}(\zeta^{test},\zeta^{true}) = \sum_{i} \frac{(N_{i}(\zeta^{test}) -  N_{i}(\zeta^{true}))^{2}}{N_{i}(\zeta^{true})}
\label{chi}
\end{equation}

If priors are included, the formula is as follows:

\begin{equation}
\chi^{2} \rightarrow \chi^{2} + \sum_{j} \frac{(\zeta_{j}^{test} - \zeta_{j}^{true})^{2}}{\sigma_{j}^{2}}
\label{chipriors}
\end{equation}	

where $\zeta^{true}$ represents the oscillation parameters that take the values from table \ref{ParametersOscillation} and $\zeta^{test}$ represents the parameters that are tested against the CBFV and assigned true VEP parameters, $N_i$ is the number of events in the $i$th bin, $\sigma_\zeta^2$ is the error in the determination of $\zeta$ and $j$ is the number of parameters with non-zero errors.

\subsection{Distorsion in the extraction of the SO parameters at DUNE} 
In this analysis we asses the possible distortions in the allowed regions of 
the SO parameters when these are obtained from neutrino oscillation data, with VEP effects inside, fitted against the pure SO formula. Considering the latter aim, we simulated DUNE data in accordance to the following parameters: $\Delta \gamma^{true} = 0 \text{, } 10^{-24} \text{, or } 2 \times 10^{-24}$, $\delta_{\mathrm{CP}}^{true}=-\pi/2$ while the remaining true values for the SO parameters are the CBFV. On the other hand, taken indeed $\Delta \gamma^{test} = 0$, we have marginalized over all SO parameters in order to find the minimum $\chi^2$.
\begin{equation}
\normalsize
    \chi^2( \theta_{13}^{test}, \delta_{\mathrm{CP}}^{test}, \Delta \gamma_{ij}^{test}=0, \theta_{13}^{true}, \delta_{\mathrm{CP}}^{true},\Delta \gamma_{ij}^{true})
    \label{chi22}
\end{equation}
The parameters that minimize the $\chi^2$ are called $\theta_{13}^{fit}$ and $\delta_{\mathrm{CP}}^{fit}$. If the contours of $\Delta \chi^2$ are analyzed on the plane $\sin^2{\theta_{13}}$ vs $\delta_{\mathrm{CP}}$, the next expression is used:
\begin{equation}
\normalsize
\begin{split}
    \Delta \chi^2=
    \chi^2( \theta_{13}^{test}, \delta_{\mathrm{CP}}^{test}, \Delta \gamma_{ij}^{test}=0, \theta_{13}^{true}, \delta_{\mathrm{CP}}^{true}, \Delta \gamma_{ij}^{true})\\ - \chi_{min}^2( \theta_{13}^{fit}, \delta_{\mathrm{CP}}^{fit}, \Delta \gamma_{ij}^{test}=0, \theta_{13}^{true}, \delta_{\mathrm{CP}}^{true}, \Delta \gamma_{ij}^{true})
\end{split}
\label{deltachi2}
\end{equation}
The same procedure described in Eqs. (\ref{chi22}) and (\ref{deltachi2})  is applied to generate the contours in the plane $\Delta m_{31}^2$ vs $\delta_{\mathrm{CP}}$.

The changes between the SO fitted allowed regions, obtained with non-null VEP data, and those regions, obtained from pure SO data with its true values fixed at the CBFV can be qualitatively 
understood   through the differences between the
VEP $\bigoplus$ SO probability, encoded in the data, and
its corresponding SO probability evaluated at the SO best fit point. Undoubtedly, and viewed at depth, the fitting of data represents the exercise of shortening the differences between the SO 
and the VEP $\bigoplus$ SO probabilities by varying (increasing or decreasing) the SO parameters in the former. Thus, it is useful to recall the approximated standard oscillation probabilities formulae engaged in our work. One is the transition oscillation channel $\nu_{\mu} \rightarrow \nu_e$ where its expression is given by:
\begin{equation}
\label{eqsome}
\begin{split}
P_{\nu_\mu \rightarrow \nu_e}^{\text{SO}} \simeq \ &C_1 s_{13}^2 + C_2 \cos \delta_{\mathrm{CP}} r s_{13} - C_3 \sin \delta_{\mathrm{CP}} r s_{13} \\ &+ C_4 r^2
\end{split}
\end{equation}
where:
\begin{equation}
\label{eqsomm}
\begin{split}
&C_1 = 4 f^2 s_{23}^2 \\
&C_2 = 8 f g \cos \Delta s_{12} c_{12} s_{23} c_{23} \\
&C_3 = 8 f g \sin \Delta s_{12} c_{12} s_{23} c_{23} \\
&C_4 = 4 g^2 s_{12}^2 c_{12}^2 c_{23}^2 \\
\end{split}
\end{equation}
All the coefficients are positive for most of the relevant energy range and the coefficients $f$ and $g$ are defined as in Eq. (\ref{defnumunue}), but without the effect of VEP.

Another relevant probability is the survival channel, $\nu_{\mu} \rightarrow \nu_{\mu}$, which has the following expression:
\begin{equation}
\begin{split}
P_{\nu_\mu \rightarrow \nu_\mu}^{\text{SO}} \simeq 1 - 4 \sin^2 \Delta s_{23}^2 c_{23}^2 + 4 \Delta \sin 2\Delta c_{12}^2 s_{23}^2 c_{23}^2 r
\end{split}
\label{survivalSO}
\end{equation}
Up to the order presented in this approximation, $\delta_{\mathrm{CP}}$ does not appear. However, for higher orders of expansion, terms proportional to $\cos \delta_{\mathrm{CP}}$ start to appear. Here, we do not present the formula up to such higher order since the size of the modifications caused by the related terms is extremely small.

\subsubsection{$\bf{U}_g = U$, $\Delta \gamma_{21} = 0$ and $\Delta \gamma_{31} \neq 0$}
\begin{figure}[h!]
	\includegraphics[scale=0.53]{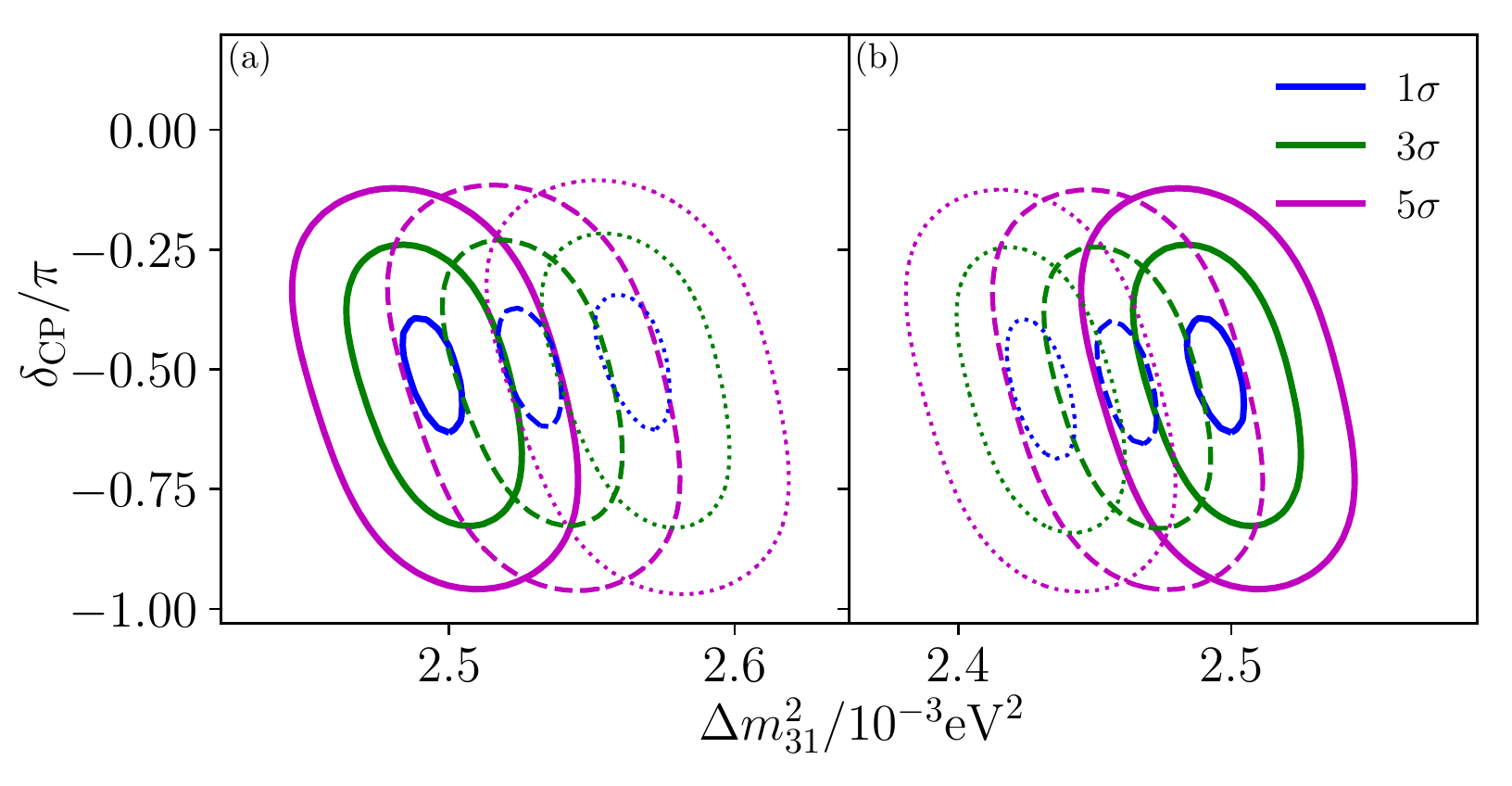}
	\caption{Scenario A/case 1. The solid lines are  $\Delta \gamma_{31}^{true}= 0$ (SO) . Figure (a) represents VEP with $ \Delta \gamma_{31}^{true}= 10^{-24}$ (dashed lines) and VEP with $ \Delta \gamma_{31}^{true}= 2 \times 10^{-24}$ (dotted lines). While in figure (b) is shown VEP with $ \Delta \gamma_{31}^{true}= -10^{-24}$ (dashed lines) and $ \Delta \gamma_{31}^{true}=- 2 \times 10^{-24}$ (dotted lines).  We consider $\delta_{\mathrm{CP}}^{true} = -\pi/2$.}
	\label{chi2s1c1dm31panel}
\end{figure}

In Fig.~\ref{chi2s1c1dm31panel} (a), the plane $\Delta m_{31}^2$ vs $\delta_{\mathrm{CP}}$ is displayed, where it is clear the shift of the 
fitted $\Delta m_{31}^2$ to higher values than the one corresponding 
to the CBFV. The shifting can be understood      
taking into account the distinct discrepancy between the VEP $\bigoplus$ SO and SO probabilities in the $\nu_{\mu} \rightarrow \nu_{\mu}$ 
channel, shown in Fig.~\ref{probs1c1mm}. As we can 
observe there, to achieve a better pairing between these probabilities it 
is required to decrease the absolute value of the SO $\nu_{\mu} \rightarrow \nu_{\mu}$ channel, which can be obtained  by increasing $\Delta m_{31}^2$ (see Eq.~(\ref{survivalSO})). Given the above explanation, when $\Delta \tilde{\gamma}_{31} < 0$, the behavior is exactly the opposite, which is observed in Fig.~\ref{chi2s1c1dm31panel} (b). The 
plane $\sin^2{\theta_{13}}$ vs $\delta_{\mathrm{CP}}$ is not shown
since the variations between allowed regions are negligible. The 
behavior of the variations  on the latter plane are 
correlated with the size of discrepancies between the VEP $\bigoplus$ SO and SO $\nu_{\mu} \rightarrow \nu_e$ probabilities, which are as a matter of fact small as shown in Fig.~\ref{probs1c1mm}.    

We have verified that if we choose, instead of VEP, any of the LV terms in the SME Hamiltonian (see section~\ref{LV}), other than the one with n = 1 energy dependency, the behavior of the allowed regions follows a similar pattern. These similarities are present in scenarios A ($\bf{U} = \bf{U}_{g}$) and B ($\bf{U} \neq \bf{U}_{g}$), throughout all the cases.

\subsubsection{$\bf U_g = U$, $\Delta \gamma_{21} \neq 0$ and $\Delta \gamma_{31} = 0$}
\begin{figure}[h!]
	\includegraphics[scale=0.53]{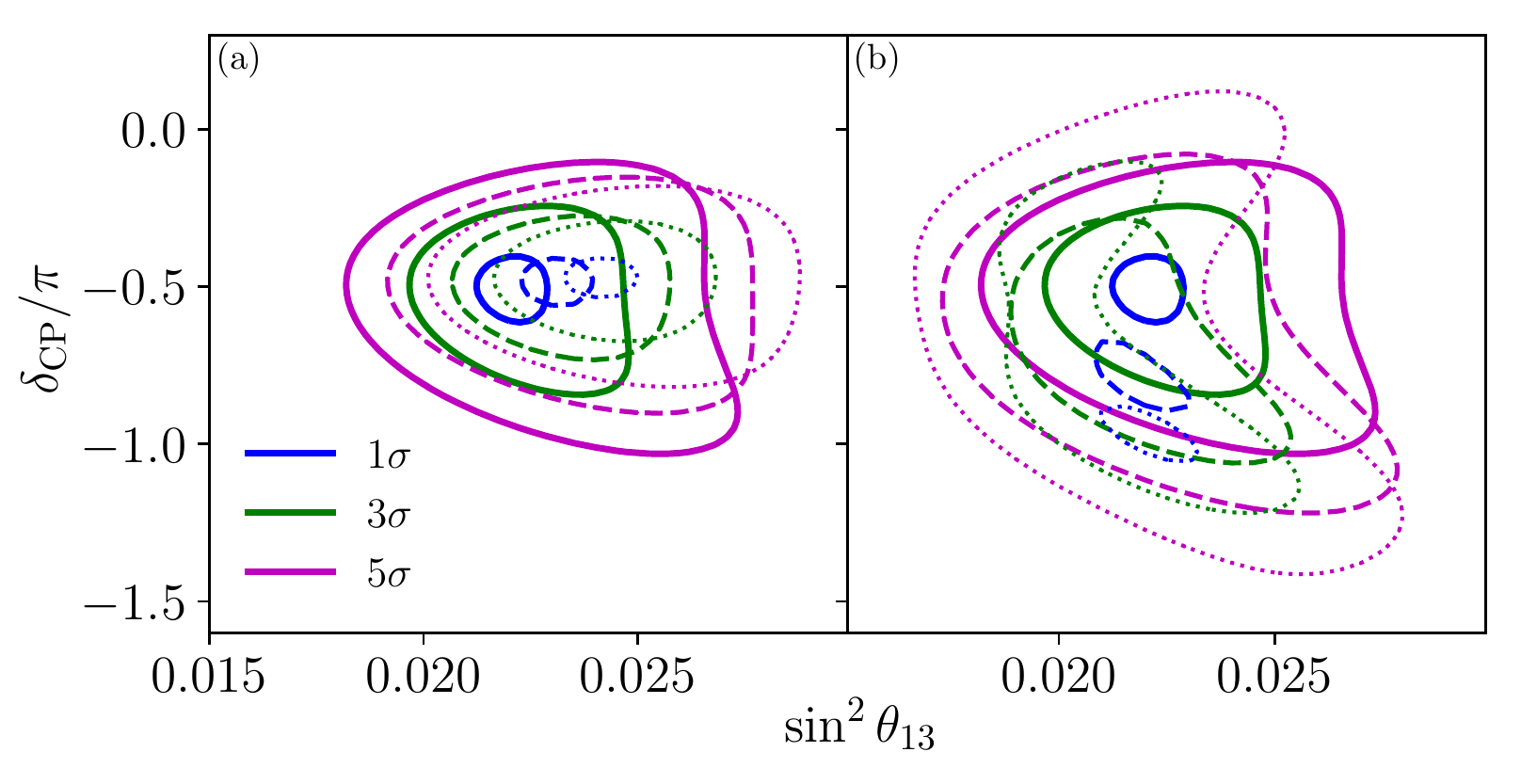}
	\caption{Scenario A/case 2.  The solid lines are  $\Delta \gamma_{21}^{true}= 0$ (SO) . Figure (a) represents VEP with $ \Delta \gamma_{21}^{true}= 10^{-24}$ (dashed lines) and VEP with $ \Delta \gamma_{21}^{true}= 2 \times 10^{-24}$ (dotted lines). While VEP with $ \Delta \gamma_{21}^{true}= -10^{-24}$ (dashed lines) and $ \Delta \gamma_{21}^{true}=- 2 \times 10^{-24}$ (dotted lines) is shown in figure (b). We consider $\delta_{\mathrm{CP}}^{true} = -\pi/2$.}
	\label{chi2s1c2panel}
\end{figure}

Contrary to the former case, in this one there are significant deviations between the allowed regions presented in the plane $\sin^2{\theta_{13}}$ vs $\delta_{\mathrm{CP}}$, as can be seen in  Fig. \ref{chi2s1c2panel}. These changes, when $\Delta \gamma_{21}>0$, are characterized by the shifting to higher values of $\sin^2{\theta_{13}}$ than the one of the SO best fit, as can be seen in Fig.~ \ref{chi2s1c2panel} (a). This shifting is explained by the need to increase  $\sin^2{\theta_{13}}$  
in order to match the SO with the VEP $\bigoplus$ SO $\nu_{\mu} \rightarrow \nu_e$ probabilities, as it is shown in Fig. \ref{probsAc2mm}. This match means to enhance the 
SO neutrino transition probability, which can be attained
by increasing the first term $C_1 s_{13}^2$, see Eq. ~(\ref{eqsome}). 
From Eq. ~(\ref{eqsome}), it is also clear that the need to decrease the SO antineutrino transition probability is satisfied through the flipped sign in term $ C_3 \sin \delta_{\mathrm{CP}} r s_{13}$. The shrinking of the allowed regions around the $\delta_{\mathrm{CP}}\sim -\pi/2$, where its effect is maximal,  happens because of the higher separation among the neutrino and antineutrino VEP $\bigoplus$ SO $\nu_\mu \rightarrow \nu_e$ probabilities than the corresponding for the SO neutrino antineutrino probability difference, evaluated at the CBFV. Therefore, in order to mimic this separation for VEP $\bigoplus$ SO neutrino-antineutrino probabilities the fitted SO probability needs to amplify the CP effects, aim which is fulfilled by choosing a narrower set of values for the $\delta_{\mathrm{CP}}$ interval around the maximal $\delta_{\mathrm{CP}} \sim -\pi/2$. When $\Delta \gamma_{21}<0$, there is a lower separation between the neutrino and antineutrino VEP $\bigoplus$ SO $\nu_\mu \rightarrow \nu_e$ probabilities and the corresponding for the SO neutrino antineutrino probability difference, at the CBFV. Then, and following the same reasoning for $\Delta \gamma_{21}>0$, but seen in opposite way, we need to adjust the fitted SO probability in order to reduce the CP effects, diminishing (increasing) the neutrino (antineutrino) SO transition channel. This can be reached through the selection of $\delta_{\mathrm{CP}}$ distant from where the maximal CP effect takes place, $\sim -\pi/2$, of the fitted SO probabilities, and, by opting for slightly smaller values of $s_{13}$ that can help modulating the reduction (rise) of the neutrino (antineutrino) transition probability magnitude (see Eq.~(\ref{eqsome})). The aforementioned behavior is totally reflected in Fig.~\ref{chi2s1c2panel} (b). In the latter figure, we can observe a misconstrued $\delta_{\mathrm{CP}}$, which is a result of how the fitted SO probabilities try to emulate the VEP effect. Finally, there is no need to display the plane $\Delta m_{31}^2$ vs $\delta_{\mathrm{CP}}$ since the discrepancies in the survival probabilities, correlated with the results in this plane, are not relevant, as seen in 
Fig. \ref{probsAc2mm}.

\subsubsection{$\bf U_g \neq U$, Texture $\theta_{13}$}

From the probabilities point of view, see Fig.~\ref{nue_sc2_case1}, this case can be seen as opposed to the preceding one. This means that for this case, $\Delta \gamma_{31} > 0 \ (\Delta \gamma_{31} < 0)$ corresponds to $\Delta \gamma_{21} < 0 \ (\Delta \gamma_{21} > 0)$ for scenario A/case 2. Therefore, the explanations for the former case could be applied to this one. On the other hand, as it can be noted in Fig.~\ref{nue_sc2_case1}, the differences between the
VEP $\bigoplus$ SO and SO $\nu_\mu \rightarrow \nu_\mu$ probabilities 
are almost null.

\subsubsection{$\bf U_g \neq U$, Texture $\theta_{12}$}

This case is equivalent to scenario A/case 2. This equivalency is rooted in the similar conduct observed in the transition probabilities, shown in
Fig.~\ref{nue_sc2_case2} and Fig.~\ref{probsAc2mm}. Hence, the arguments used for explaining the allowed regions behavior for scenario A/case 2 are totally suitable to be applied to this case.

\subsubsection{$\bf U_g \neq U$, Texture $\theta_{23}$}

As pointed out in sections~\ref{scenarioBsubcase3a} and~\ref{scenarioBsubcase3b} only in the 
$\nu_\mu \rightarrow \nu_\mu$ channel the 
discrepancies between the VEP $\bigoplus$ SO and the SO are observable (evaluated at the CBFV). Therefore, the plane $\Delta m_{31}^2$ vs $\delta_{\mathrm{CP}}$ is the appropriate parameter space region, where the impact of these differences can be revealed. Scenario B/texture $\theta_{23}$-a, $\Delta \gamma_{21} = 0$ and $\Delta \gamma_{31} \neq 0$, exhibits a quite similar behavior to that shown in Fig.~\ref{chi2s1c1dm31panel} for scenario A/case 1. Scenario B/texture $\theta_{23}$-b, $\Delta \gamma_{31} = 0, \Delta \gamma_{21}>0 \ (\Delta \gamma_{21}<0)$ corresponds to $\Delta \gamma_{31}<0 \ (\Delta \gamma_{31}>0)$ for scenario A/case 1.  Both tendencies in Fig.~\ref{chi2s1c1dm31panel} (a) and (b) are in agreement to what is expected from the probabilities displayed in Fig.~\ref{numu_sc2_case3}. For texture $\theta_{23}$-a ($\theta_{23}$-b), the fitted SO probability has to lessen (augment) its value to match with the VEP $\bigoplus$ SO, which means to increase (decrease) $\Delta m_{31}^2$, as can be checked in Eq.~(\ref{survivalSO}).

\subsection{VEP Sensitivity limits}

We analyze the sensitivity of DUNE to VEP parameters generating a pure standard oscillation simulated data, fixing the following true values: $\Delta \gamma^{true}=0$, and a given value of $\delta_{\mathrm{CP}}^{true}$, marginalizing over the remaining standard oscillation parameters. 
\begin{equation}
    \chi^2=\chi^2 \left( \Delta \gamma^{test},\delta_{\mathrm{CP}}^{true},\Delta \gamma^{true}=0\right)
    \label{chisensitivity}
\end{equation}
The $\Delta \gamma^{test}$ is the test parameter paying attention 
that $\Delta \gamma^{true} (\Delta \gamma^{test})$ either would take the value of $\Delta \gamma_{31}^{true}(\Delta \gamma_{31}^{test}) $ or $\Delta \gamma_{21}^{true}(\Delta \gamma_{21}^{test})$ depending on the case to be studied.

\subsubsection{Scenario A}

\begin{figure}
    \includegraphics[scale=0.53]{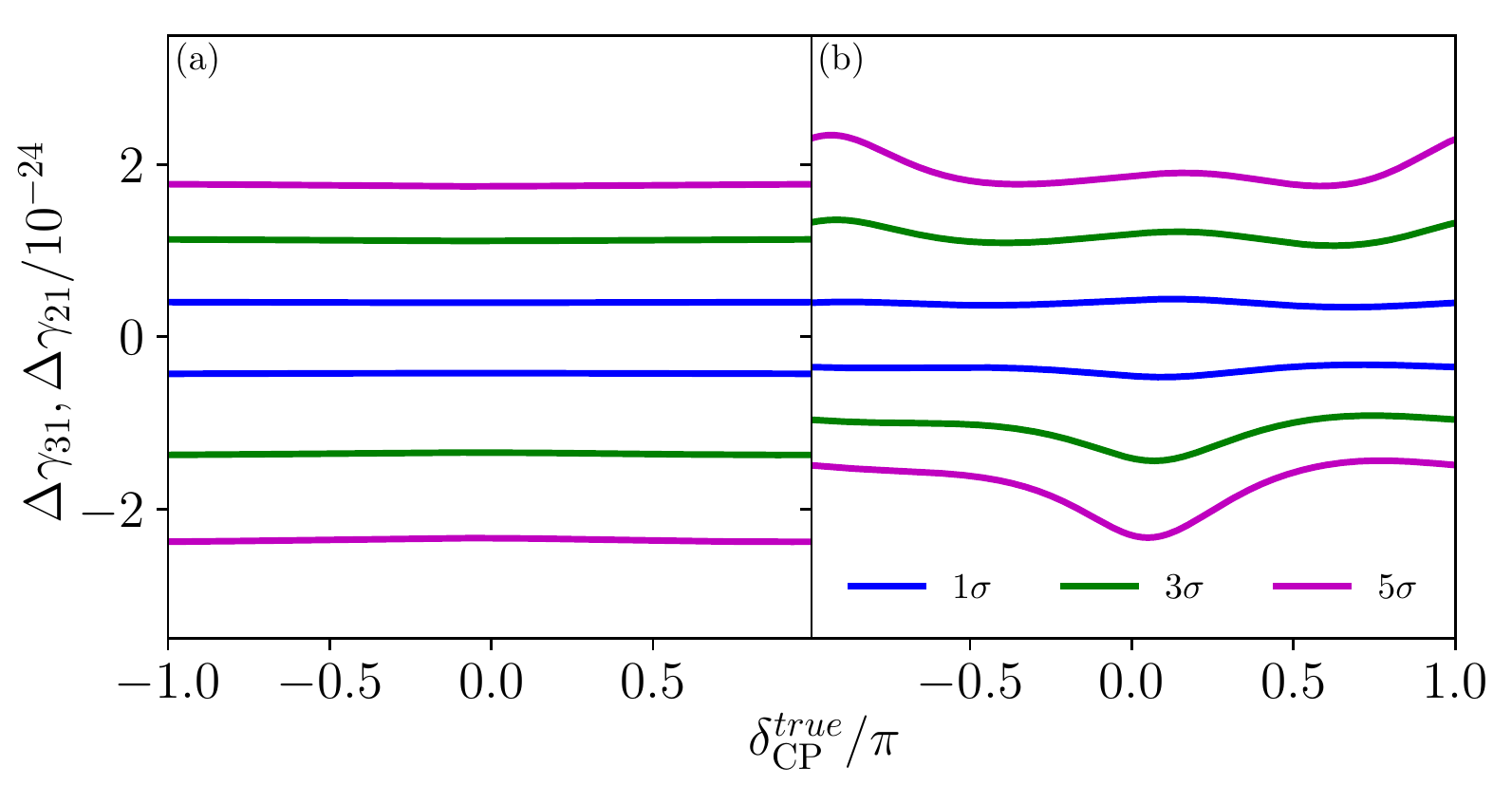}
    \caption{Sensitivity to VEP considering scenario A/case 1 (a) and  case 2 (b), depending on $\delta_{\mathrm{CP}}^{true}$. }
    \label{chilimits}
\end{figure}

 In  Fig. \ref{chilimits} it is displayed the sensitivity to the VEP parameter for the different cases of scenario A. For case 1, the sensitivity to $\Delta \gamma_{31}$ is given by $\left[0.4, 1.1, 1.8 \right]\times 10^{-24}$ and $-\left[0.4, 1.4, 2.4 \right]\times 10^{-24}$ at the $1\sigma$, $3\sigma$, and $5\sigma$ levels, respectively. In this plot we can see that the sensitivity to $\Delta \gamma_{31}$  is almost constant irregardless the value of $\delta_{\mathrm{CP}}$. The latter can be inferred from the probabilities given in Eqs.(\ref{eqs1c1me}) and (\ref{nunus1c1me}), where $\delta_{\mathrm{CP}}$ is not appearing, unless up to the perturbation order that we present in these formulae. When we consider negative values of $\Delta \gamma_{31}$, the formula predicts the same correction, which implies a same constant behavior, and rather similar values for the sensitivity, as the positive case. This can be seen in Fig. \ref{chilimits}.
 
\begin{figure*}
    \includegraphics[scale=0.39]{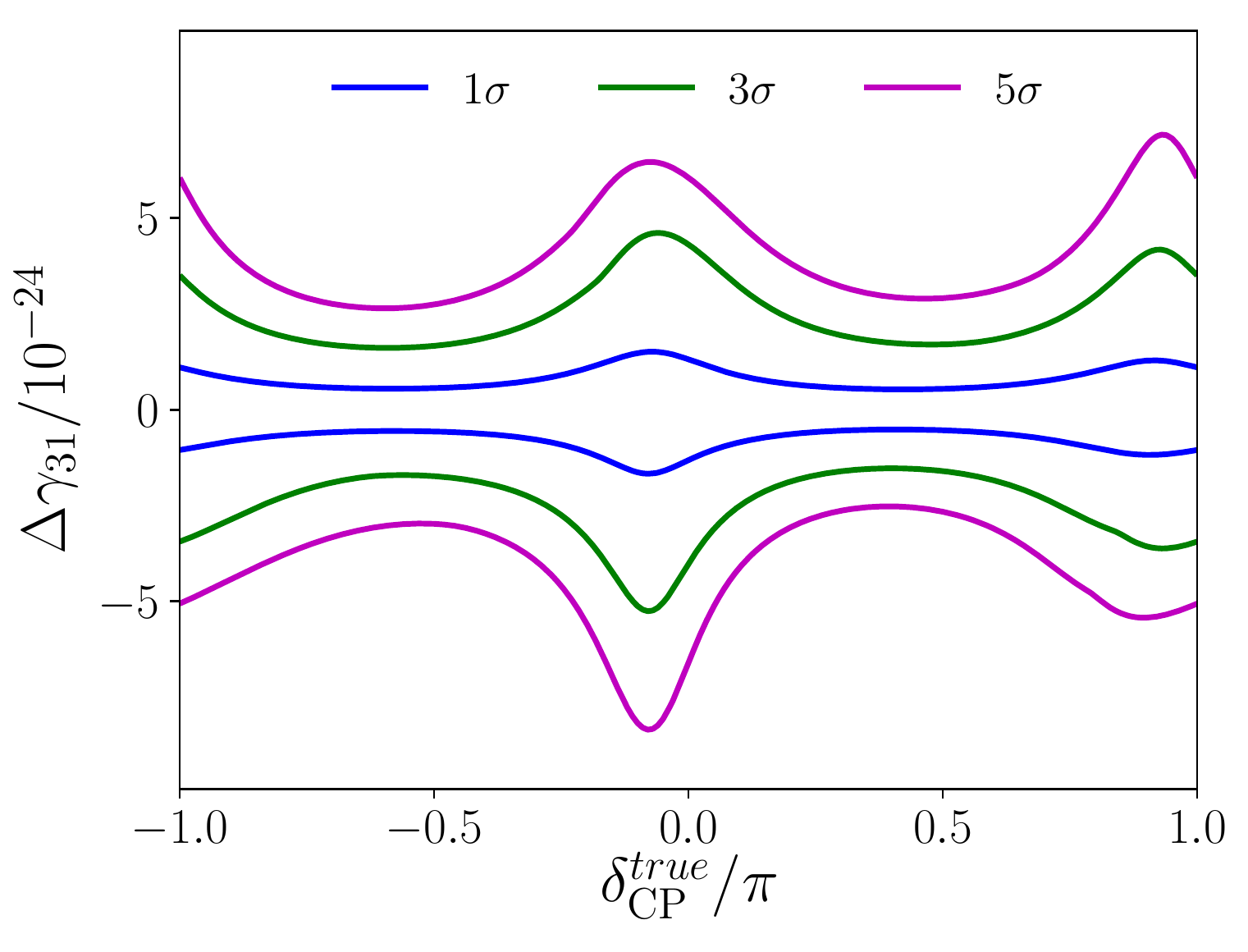}\hspace{-0.21cm}
   \includegraphics[scale=0.39]{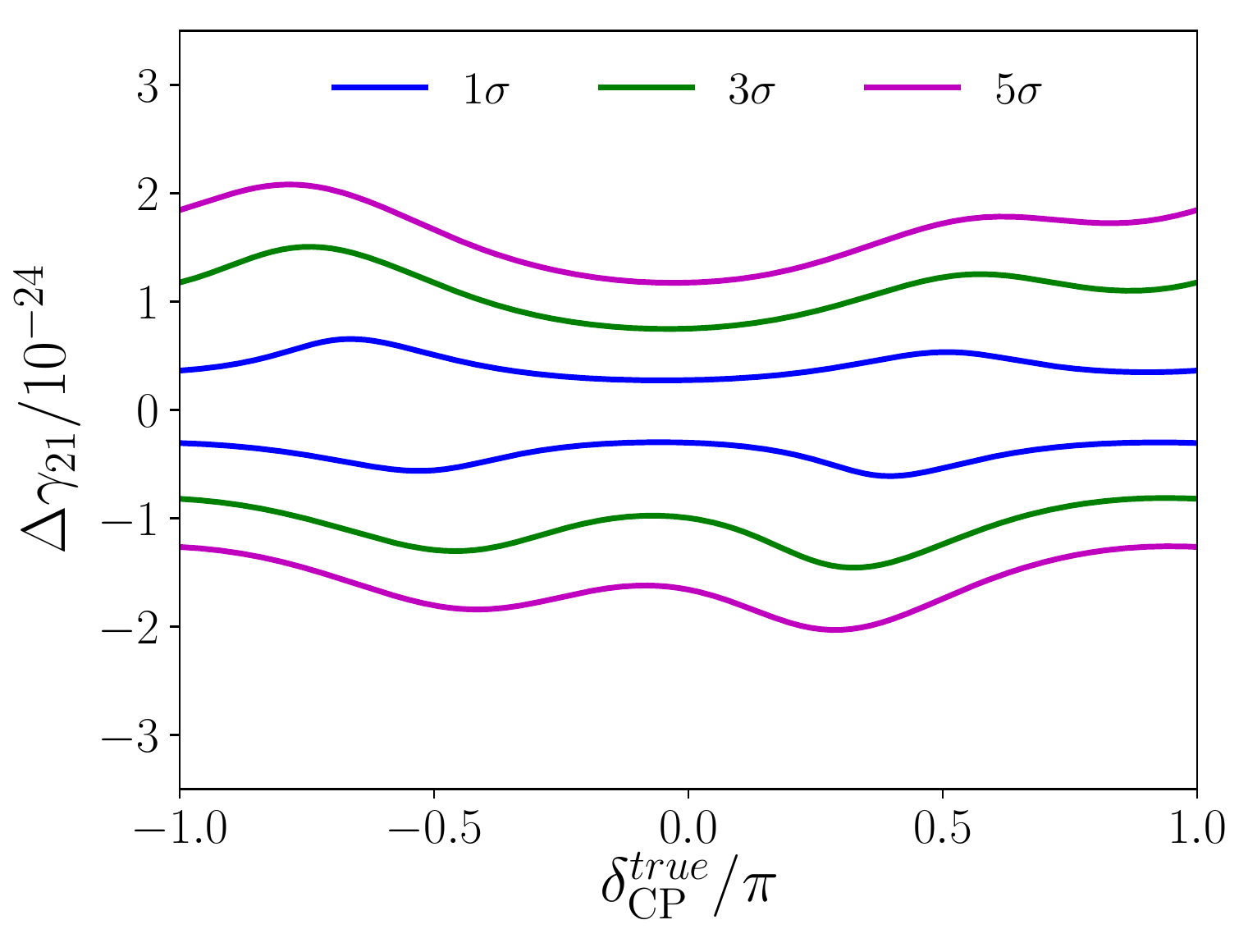}\hspace{-0.21cm}
   \includegraphics[scale=0.39]{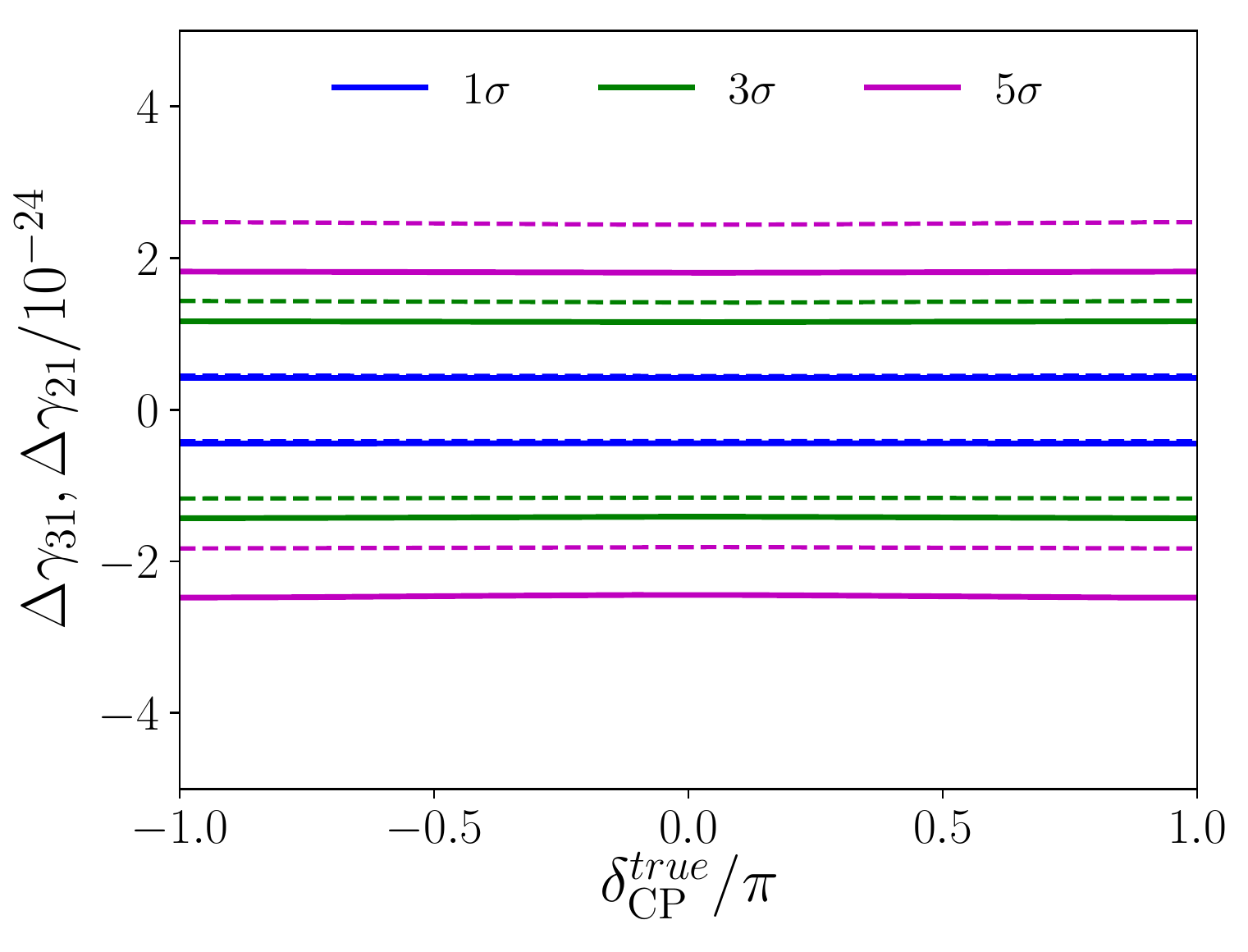}
    \caption{Sensitivity to VEP considering scenario B/textures $\theta_{13}$ (left), $\theta_{12}$ (center) and $\theta_{23}$ (right) depending on $\delta_{\mathrm{CP}}^{true}$. In the plot on the right, the solid and dashed lines represent the sub-cases \textit{a} and \textit{b} respectively. We consider $\theta_{12}^{g}$, $\theta_{23}^{g}$ and $\theta_{13}^{g}$ equal to $\pi/4$.}
    \label{chilimits2}
\end{figure*}

In this figure a plot for case 2 is shown, as well. For this case, the sensitivity to $\Delta \gamma_{21}$ for its positive values is $\left[ 0.3 \ \textendash \ 0.4,1.1 \ \textendash \ 1.4,1.8 \ \textendash \ 2.4 \right]\times 10^{-24}$ and for its negative values is $-\left[0.3 \ \textendash \ 0.5,0.9 \ \textendash \ 1.4,1.4 \ \textendash \ 2.3 \right]\times 10^{-24}$ at the $1\sigma$, $3\sigma$, and $5\sigma$ levels.  As it can be seen from Fig.~\ref{probsAc2mm}, the highest discrepancies between VEP $\bigoplus$ SO and pure SO are present in the $\nu_\mu \rightarrow \nu_e$ transition channel. Consequently, it should be expected that the shape of the curve of the sensitivity is affected, at some degree, by the transition channel. Therefore, for getting a qualitative understanding of this shape we use the analytical expression of the $\nu_\mu \rightarrow \nu_e$ transition channel. In particular, the two lowest order perturbative (most relevant) terms in Eq.~(\ref{eqs1c2me}) can be grouped into a single term proportional to $\cos (\Delta + \delta_{\mathrm{CP}})$. Fixing the neutrino energy at $2.5 \ \mathrm{GeV}$ (the mean energy at DUNE), for which $\Delta$ is close to $ 0.5\pi$, it is possible to have a rough idea about the location of the maximum and minimum sensitivities. Then, if $\Delta$ is close to $ 0.5\pi$, it is expected that the maximum sensitivity points are located in values of  $\delta_{\mathrm{CP}}$ in the vicinity of  
$-0.5 \pi$ and $0.5 \pi$. This is what we observe for positive values of $\Delta \gamma_{21}$. Before we continue, it is convenient to point out that maximum sensitivity points correspond to the lowest deflections of the VEP $\bigoplus$ SO -probability respect the SO one. On the other hand, minimum sensitivity is obtained for values of $\delta_{\mathrm{CP}}$ at the vicinity of $0$, $\pi$ and $-\pi$. For negative values of $\Delta \gamma_{21}$, minimum sensitivity for $\delta_{\mathrm{CP}}$ close to $0$ still survives. However, the other minima and maxima are erased because of the influence of the terms following the first and second ones in the correction.

\subsubsection{Scenario B}

In the same way, Fig. \ref{chilimits2} shows the sensitivity to the new parameters for textures $\theta_{13}$, $\theta_{12}$ and $\theta_{23}$ of scenario B. First we focus on texture $\theta_{13}$ and texture $\theta_{12}$. For texture $\theta_{13}$, the sensitivity to $\Delta \gamma_{31}$ is given by $[0.5 \ \textendash \ 1.5, 1.6 \ \textendash \ 4.6, 2.6 \ \textendash \ 7.2]\times 10^{-24}$ for the positive values and $-[0.5 \ \textendash \ 1.7, 1.5 \ \textendash \ 5.3, 2.5 \ \textendash \ 8.4]\times 10^{-24}$ for the negative ones at the $1\sigma$, $3\sigma$, and $5\sigma$ levels respectively. For texture $\theta_{12}$ of the same scenario, the sensitivity to $\Delta \gamma_{21}$ is given by $[0.3 \ \textendash \ 0.7, 0.7 \ \textendash \ 1.5, 1.2 \ \textendash \ 2.1]\times 10^{-24}$ and $-[0.3 \ \textendash \ 0.6, 0.8 \ \textendash \ 1.5, 1.3 \ \textendash \ 2]\times 10^{-24}$ at the $1\sigma$, $3\sigma$, and $5\sigma$ levels respectively.

The sensitivity behavior for theses textures, $\theta_{13}$ and $\theta_{12}$, is almost absolutely dominated by the 
 $\nu_\mu \rightarrow \nu_e$ transition channel, given that only in this channel there are (observable) discrepancies between VEP $\bigoplus$ SO and pure SO (see Figs.~\ref{nue_sc2_case1} and ~\ref{nue_sc2_case2}). In particular, it is possible to get a feeling of the approximated position of the maximum and minimum sensitivity points analyzing the first two terms in the transition probabilities for both textures. These two terms are proportional to $ C_1 \cos \delta_{\mathrm{CP}} \pm  C_2  \sin \delta_{\mathrm{CP}}$. Then, when $C_1 < C_2  (C_1 > C_2) $  the maximum (minimum) sensitivity in $\delta_{\mathrm{CP}}$ is located in the neighborhood of $0.5 \pi$ and $-0.5\pi$ ($0$, $\pi$,  and $-\pi$) for texture $\theta_{13}$ (textures $\theta_{12}$). In the minimum (maximum) sensitivity point is where  
the lowest (highest) discrepancies between VEP $\bigoplus$ SO and pure SO are found. For both signs of $\Delta \gamma$ the behavior is similar, unless, of course, some shifts due to the influence of the other terms.

Fig. \ref{chilimits2} presents the sensitivity to  $\Delta \gamma_{31}$ and  $\Delta \gamma_{21}$ in the context of scenario B, texture $\theta_{23}$ and sub-cases \textit{a} and \textit{b} respectively. Thus, the sensitivity to $\Delta \gamma_{31}$ $(\Delta \gamma_{21})$ is given by $[0.4, 1.2, 1.8]\times 10^{-24}$ and $-[0.4, 1.4, 2.5]\times 10^{-24}$ ($[0.4, 1.4, 2.5]\times 10^{-24}$ and $-[0.4, 1.2, 1.8]\times 10^{-24}$) at the $1\sigma$, $3\sigma$, and $5\sigma$ levels respectively. It is important to note that in both sub-cases the dependence on $\delta_{\mathrm{CP}}$ is negligible, since, there are only deviations from SO in the $\nu_\mu \rightarrow \nu_\mu$ survival channel. For sub-cases \textit{a} and \textit{b}, there are no VEP-related terms in the transition probability $\nu_{\mu} \rightarrow \nu_e$ up to the level of the developed perturbation order. On the other hand, sub-case \textit{a} deflects from the SO case more visibly than sub-case \textit{b}. That is why the former has higher sensitivity than the latter. It is good to mention that the aforementioned situation cannot be easily noted in the corresponding probability plots (see Fig.~\ref{numu_sc2_case3}). In addition, there is a symmetric behavior for both signs of  $\Delta \gamma_{ij}$.

\subsection{Lorentz Violation Sensitivity Limits}
As we have pointed out our VEP prescription can be reapplied to test the different isotropic Lorentz violating terms of the SME Hamiltonian with their respectives energy dependencies, as discussed in section~\ref{LV}. Here we have set up different limits imposed on each of the aforementioned terms, in the context of DUNE, working with them in individual manner. Since this is an indirect result of this manuscript, we only present them on table \ref{limitsLIV}. As similar works can be found in \cite{Jurkovich:2018rif,Barenboim:2018ctx}.

Table \ref{limitsLIV} presents the sensitivity of DUNE experiment to LV. It can be seen that scenario B/texture $\theta_{12}$ shows the greatest constraint to the parameter $\Delta \gamma_{21}$ for almost all $n$. In the meantime, scenario B/texture $\theta_{13}$ presents precisely the opposite for constraining  $\Delta \gamma_{31}$. This is exactly the same pattern found for VEP, whence the explanation is the same. Therefore, scenario B/texture $\theta_{13}$ is sensitive to higher $\Delta \gamma_{31}$ values, while scenario B/texture $\theta_{12}$ is sensitive to lower $\Delta \gamma_{21}$ values. 

\begin{table*}[htb]
\begin{center}
\begin{tabular}{|c|c|c|c|c|}
\hline
\parbox[c][][c]{7em}{ \small{Scenario/ \\ case, texture} } & $n=0$ & $n=1$ & $n=2$ & $n=3$	\\
\hline 
A/1 & $\Delta \gamma_{31}\times 10^{-23}   \mathrm{GeV}$ & $\Delta \gamma_{31}\times  10^{-24}$ & $ \Delta \gamma_{31}\times 10^{-25}   \mathrm{GeV}^{-1}$ & $\Delta \gamma_{31}\times  10^{-26}   \mathrm{GeV}^{-2}$ \\
\hline 

			1$\sigma$ &\begin{tabular}{c c}
                   0.5 \ \ \ & \ \ \  -0.5 \\
                 \end{tabular}			
                &\begin{tabular}{c c}
                   0.8  \ \ \ & \ \ \  -0.9 \\
                 \end{tabular} & \begin{tabular}{c c}
                   0.6 \ \ \ & \ \ \ -0.8 \\
                 \end{tabular} & \begin{tabular}{c c}
                   0.4 \ \ \ & \ \ \ -0.5 \\
                 \end{tabular} \\
			3$\sigma$ & \begin{tabular}{c c}
                   1.4 \ \ \ & \ \ \ -1.3 \\
                 \end{tabular}&\begin{tabular}{c c}
                   2.2 \ \ \ & \ \ \ -2.7 \\
                 \end{tabular} & \begin{tabular}{c c}
                   1.6 \ \ \ & \ \ \ -3.3 \\
                 \end{tabular} & \begin{tabular}{c c}
                   0.9 \ \  \ & \ \ \ -2.3 \\
                 \end{tabular}\\
			5$\sigma$ & \begin{tabular}{c c}
                   2.3 \ \ \ & \ \ \ -2.2 \\
                 \end{tabular}&\begin{tabular}{c c}
                   3.5  \ \ \ & \ \ \ -4.7 \\
                 \end{tabular} & \begin{tabular}{c c}
                   2.4 \ \ \ & \ \ \ -5.3 \\
                 \end{tabular} & \begin{tabular}{c c}
                   1.4 \ \ \ & \ \ \ -3.2 \\
                 \end{tabular} \\ 
			\hline
A/2 & $ \Delta \gamma_{21} \times 10^{-23}   \mathrm{GeV}$ & $ \Delta \gamma_{21} \times 10^{-24}$ & $ \Delta \gamma_{21} \times 10^{-25}   \mathrm{GeV}^{-1}$ & $ \Delta \gamma_{21} \times 10^{-26}   \mathrm{GeV}^{-2}$ \\
\hline 
			1$\sigma$ &\begin{tabular}{c c}
                   [0.2 $-$ 0.4] \  & \  -[0.2 $-$ 0.4] \\
                 \end{tabular}			
                &\begin{tabular}{c c}
                   [0.7 $-$ 0.9]  \ &  \ -[0.6 $-$ 0.9] \\
                 \end{tabular} & \begin{tabular}{c c}
                   [0.8 $-$ 1.1]  \ & \  -[0.6 $-$ 1.0] \\
                 \end{tabular} & \begin{tabular}{c c}
                   [0.5 $-$ 0.8] \  &  \ -[0.3 $-$ 0.6] \\
                 \end{tabular} \\
			3$\sigma$ & \begin{tabular}{c c}
                   [0.7 $-$ 1.3] \  &  \ -[0.7 $-$ 1.3] \\
                 \end{tabular}&\begin{tabular}{c c}
                   [2.1 $-$ 2.7] \  & \  -[1.8 $-$ 2.9] \\
                 \end{tabular} & \begin{tabular}{c c}
                   [2.1 $-$ 3.0] \  & \  -[1.4 $-$ 2.3] \\
                 \end{tabular} & \begin{tabular}{c c}
                   [1.3 $-$ 1.8] \ &  \ -[0.8 $-$ 1.3] \\
                 \end{tabular}\\
			5$\sigma$ & \begin{tabular}{c c}
                   [1.2 $-$ 2.1]  \ & \  -[1.3 $-$ 2.3] \\
                 \end{tabular}&\begin{tabular}{c c}
                   [3.5 $-$ 4.7]  \ & \  -[2.9 $-$ 4.7] \\
                 \end{tabular} & \begin{tabular}{c c}
                   [3.2 $-$ 4.2] \  & \  -[2.1 $-$ 3.3] \\
                 \end{tabular} & \begin{tabular}{c c}
                   [2.0 $-$ 2.4]  \ &  \ -[1.2 $-$ 1.9] \\
                 \end{tabular} \\ 
			\hline
B/$\theta_{13}$ & $ \Delta \gamma_{31} \times 10^{-23}   \mathrm{GeV}$ & $ \Delta \gamma_{31} \times 10^{-24}$ & $ \Delta \gamma_{31} \times 10^{-25}   \mathrm{GeV}^{-1}$ & $ \Delta \gamma_{31} \times 10^{-26}   \mathrm{GeV}^{-2}$ \\
\hline 
			1$\sigma$ &\begin{tabular}{c c}
                   [0.3 $-$ 0.8]  \ &  \ -[0.3 $-$ 0.9] \\
                 \end{tabular}			
                &\begin{tabular}{c c}
                   [1.1 $-$ 3.0] \  &  \ -[1.0 $-$ 3.3] \\
                 \end{tabular} & \begin{tabular}{c c}
                   [2.7 $-$ 5.5]  \ & \  -[2.6 $-$ 5.4] \\
                 \end{tabular} & \begin{tabular}{c c}
                   [2.4 $-$ 3.6] \  &  \ -[2.3 $-$ 3.4] \\
                 \end{tabular} \\
			3$\sigma$ & \begin{tabular}{c c}
                   [1.0 $-$ 2.8] \  & \  -[0.9 $-$ 3.0] \\
                 \end{tabular}&\begin{tabular}{c c}
                   [3.2 $-$ 9.2]   & \  -[3.1 $-$ 10.5] \\
                 \end{tabular} & \begin{tabular}{c c}
                   [6.7 $-$ 10.9]   &   \ -[6.5 $-$ 9.9] \\
                 \end{tabular} & \begin{tabular}{c c}
                   [5.1 $-$ 7.3]  \ &   \ -[5.1 $-$ 6.8] \\
                 \end{tabular}\\
			5$\sigma$ & \begin{tabular}{c c}
                   [1.6 $-$ 4.1] \  &  \ -[1.5 $-$ 7.6] \\
                 \end{tabular}&\begin{tabular}{c c}
                   [5.3 $-$ 14.3]   &   -[5.0 $-$ 16.7] \\
                 \end{tabular} & \begin{tabular}{c c}
                   [9.9 $-$ 46.1]   &   -[9.6 $-$ 16.2] \\
                 \end{tabular} & \begin{tabular}{c c}
                   [7.6 $-$ 83.6]   &   -[7.2 $-$ 70.0] \\
                 \end{tabular} \\ 
			\hline
B/$\theta_{12}$ & $ \Delta \gamma_{21} \times 10^{-23}   \mathrm{GeV}$ & $ \Delta \gamma_{21} \times 10^{-24}$ & $ \Delta \gamma_{21} \times 10^{-25}   \mathrm{GeV}^{-1}$ & $ \Delta \gamma_{21} \times 10^{-26}   \mathrm{GeV}^{-2}$ \\
\hline 
			1$\sigma$ &\begin{tabular}{c c}
                   [0.3 $-$ 0.4]  \ & \  -[0.3 $-$ 0.4] \\
                 \end{tabular}			
                &\begin{tabular}{c c}
                   [0.5 $-$ 1.3] \  & \  -[0.6 $-$ 1.2] \\
                 \end{tabular} & \begin{tabular}{c c}
                   [0.5 $-$ 1.3] \  &  \ -[0.6 $-$ 1.3] \\
                 \end{tabular} & \begin{tabular}{c c}
                   [0.3 $-$ 0.7] \  &  \ -[0.3 $-$ 0.7] \\
                 \end{tabular} \\
			3$\sigma$ & \begin{tabular}{c c}
                   [0.9 $-$ 1.2] \  & \  -[0.8 $-$ 1.2] \\
                 \end{tabular}&\begin{tabular}{c c}
                   [1.5 $-$ 3.0] \  &  \ -[1.6 $-$ 2.9] \\
                 \end{tabular} & \begin{tabular}{c c}
                   [1.2 $-$ 2.2] \  &  \ -[1.3 $-$ 2.2] \\
                 \end{tabular} & \begin{tabular}{c c}
                   [0.7 $-$ 1.2] \  &  \ -[0.8 $-$ 1.3] \\
                 \end{tabular}\\
			5$\sigma$ & \begin{tabular}{c c}
                   [1.4 $-$ 2.1] \  & \  -[1.4 $-$ 2.0] \\
                 \end{tabular}&\begin{tabular}{c c}
                   [2.3 $-$ 4.2] \  & \  -[2.5 $-$ 4.1] \\
                 \end{tabular} & \begin{tabular}{c c}
                   [1.8 $-$ 2.9] \  & \  -[1.9 $-$ 2.9] \\
                 \end{tabular} & \begin{tabular}{c c}
                   [1.1 $-$ 1.6] \  &  \ -[1.1 $-$ 1.7] \\
                 \end{tabular} \\ 
			\hline
B/$\theta_{23}$-a & $\Delta \gamma_{31}\times 10^{-23}   \mathrm{GeV}$ & $\Delta \gamma_{31}\times  10^{-24}$ & $ \Delta \gamma_{31}\times 10^{-25}   \mathrm{GeV}^{-1}$ & $\Delta \gamma_{31}\times  10^{-26}   \mathrm{GeV}^{-2}$ \\
\hline 

			1$\sigma$ &\begin{tabular}{c c}
                   0.5 \ \ \ & \ \ \  -0.5 \\
                 \end{tabular}			
                &\begin{tabular}{c c}
                   0.8  \ \ \ & \ \ \  -0.9 \\
                 \end{tabular} & \begin{tabular}{c c}
                   0.6 \ \ \ &  \ \ \ -0.8 \\
                 \end{tabular} & \begin{tabular}{c c}
                   0.4 \ \ \ & \ \ \ -0.5 \\
                 \end{tabular} \\
			3$\sigma$ & \begin{tabular}{c c}
                   1.4 \ \ \ & \  \ \ -1.4 \\
                 \end{tabular}&\begin{tabular}{c c}
                   2.3 \ \ \ & \ \ \ -2.8 \\
                 \end{tabular} & \begin{tabular}{c c}
                   1.7 \ \ \ & \ \ \ -3.3 \\
                 \end{tabular} & \begin{tabular}{c c}
                   1.0 \ \ \ & \ \ \ -2.2 \\
                 \end{tabular}\\
			5$\sigma$ & \begin{tabular}{c c}
                   2.3 \ \ \ & \ \ \ -2.3 \\
                 \end{tabular}&\begin{tabular}{c c}
                   3.6  \ \ \ & \ \ \ -4.9 \\
                 \end{tabular} & \begin{tabular}{c c}
                   2.5 \ \ \ & \ \ \ -5.5 \\
                 \end{tabular} & \begin{tabular}{c c}
                   1.4 \ \ \ & \ \ \ -3.2 \\
                 \end{tabular} \\ 
			\hline	
B/$\theta_{23}$-b & $\Delta \gamma_{21}\times 10^{-23}   \mathrm{GeV}$ & $\Delta \gamma_{21}\times  10^{-24}$ & $ \Delta \gamma_{21}\times 10^{-25}   \mathrm{GeV}^{-1}$ & $\Delta \gamma_{21}\times  10^{-26}   \mathrm{GeV}^{-2}$ \\
\hline 

			1$\sigma$ &\begin{tabular}{c c}
                   0.5 \ \ \ & \ \ \ -0.5 \\
                 \end{tabular}			
                &\begin{tabular}{c c}
                   0.9  \ \ \ & \ \ \ -0.8 \\
                 \end{tabular} & \begin{tabular}{c c}
                   0.8 \ \ \ &  \ \ \ -0.7 \\
                 \end{tabular} & \begin{tabular}{c c}
                   0.5 \ \  \ & \  \ \ -0.4 \\
                 \end{tabular} \\
			3$\sigma$ & \begin{tabular}{c c}
                   1.4 \ \ \ &  \ \ \ -1.4 \\
                 \end{tabular}&\begin{tabular}{c c}
                   2.8 \ \ \ & \ \ \ -2.3 \\
                 \end{tabular} & \begin{tabular}{c c}
                   3.3 \ \ \ & \ \ \ -1.7 \\
                 \end{tabular} & \begin{tabular}{c c}
                   2.3 \ \ \ & \ \ \ -1.0 \\
                 \end{tabular}\\
			5$\sigma$ & \begin{tabular}{c c}
                   2.3 \ \ \ & \ \ \ -2.3 \\
                 \end{tabular}&\begin{tabular}{c c}
                   4.9  \ \ \ & \ \ \ -3.6 \\
                 \end{tabular} & \begin{tabular}{c c}
                   5.3 \ \ \ & \ \ \ -2.5 \\
                 \end{tabular} & \begin{tabular}{c c}
                   3.2 \ \ \ & \ \ \ -1.4 \\
                 \end{tabular} \\ 
			\hline											
		\end{tabular}
		\caption{Limits for Lorentz violation. For the scenario B,  $\theta_{12}^{g}$, $\theta_{23}^{g}$ and $\theta_{13}^{g}$ are considered equal to $\pi/4$.}
		\label{limitsLIV}
\end{center}\end{table*}

\subsection{CP Violation and Mass Hierarchy}

\subsubsection{CP Violation Sensitivity}

This section discusses the effect of VEP on CP violation sensitivity at DUNE experiment. To refer to DUNE sensitivity to CP violation, the definition shown in \cite{Acciarri:2015uup,Carpio:2018gum} are taken into account.

\begin{equation}
\begin{split}
\Delta \chi_{\mathrm{CP}}^2 =& \mathrm{Min} [ \Delta \chi^2 (\delta^{test}=0,\Delta \gamma^{test}=0,\delta^{true},\Delta \gamma^{true}), \\
 & \Delta \chi^2 (\delta^{test}=\pi,\Delta \gamma^{test}=0,\delta^{true},\Delta \gamma^{true})]
\end{split}
\end{equation}

To calculate $\Delta\chi_{CP}^2$, $\delta_{\mathrm{CP}}$ and $\Delta \gamma$ are set as fixed  while it is marginalized over the rest of the parameters.
 The CP violation sensitivity is studied by fitting the data as SO and considering VEP as an unknown but existing effect. In most cases it is observed an increase in the significance level to reject the null hypothesis depending on $\delta_{\mathrm{CP}}^{true}$. However, some cases show a decrease of this significance level for certain values of $\delta_{\mathrm{CP}}^{true}$, all with respect to SO. This way of analysis is very important to study the consequences of omitting an existing VEP scenario in nature in our theoretical framework.

\begin{figure}[h!]
	\includegraphics[scale=0.53]{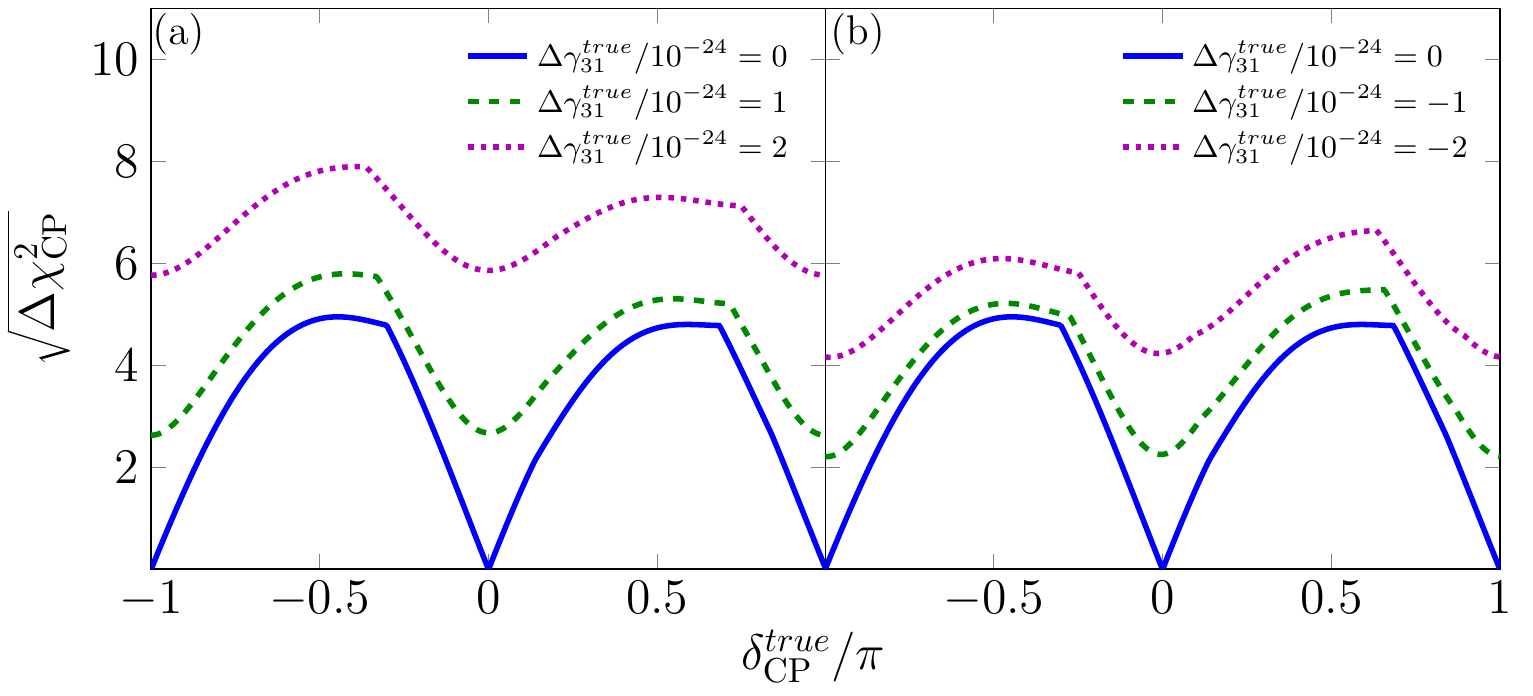}
	\caption{ CP Violation sensitivity for scenario A/case 1.}
	\label{CPV_sensi_scA_cs1_panel}
\end{figure}

In Fig. \ref{CPV_sensi_scA_cs1_panel}, scenario A/case 1, an increase in the significance level to reject the null hypothesis can be observed even when $\delta_{\mathrm{CP}}^{true}=0,\pm \pi$, generating a fake CPV. This is because there is a relatively constant increment on sensitivity and is a reflection of the $\delta_\mathrm{CP}$-independent discrepancy between the VEP $\bigoplus$ SO and SO in the $\nu_\mu \ ( \text{and } \bar{\nu}_\mu)$ disappearance probabilities for scenario A/case 1 (see Eq.~(\ref{nunus1c1me}). The increase of the number of events for the $\Delta\gamma<0$ reduces the $\sqrt{\Delta \chi^{2}}$ making it harder to achieve similar values of sensitivity to those obtained for the $\Delta\gamma>0$ case. These results are qualitatively similar to those shown in scenario B/texture $\theta_{23}$-a. Additionally, scenario B/texture $\theta_{23}$-b  $\Delta \gamma_{21}>0 \ (\Delta \gamma_{21}<0)$ corresponds to $\Delta \gamma_{31}<0 \ (\Delta \gamma_{31}>0)$ for scenario A/case 1.

\begin{figure}[h!]
	\includegraphics[scale=0.55]{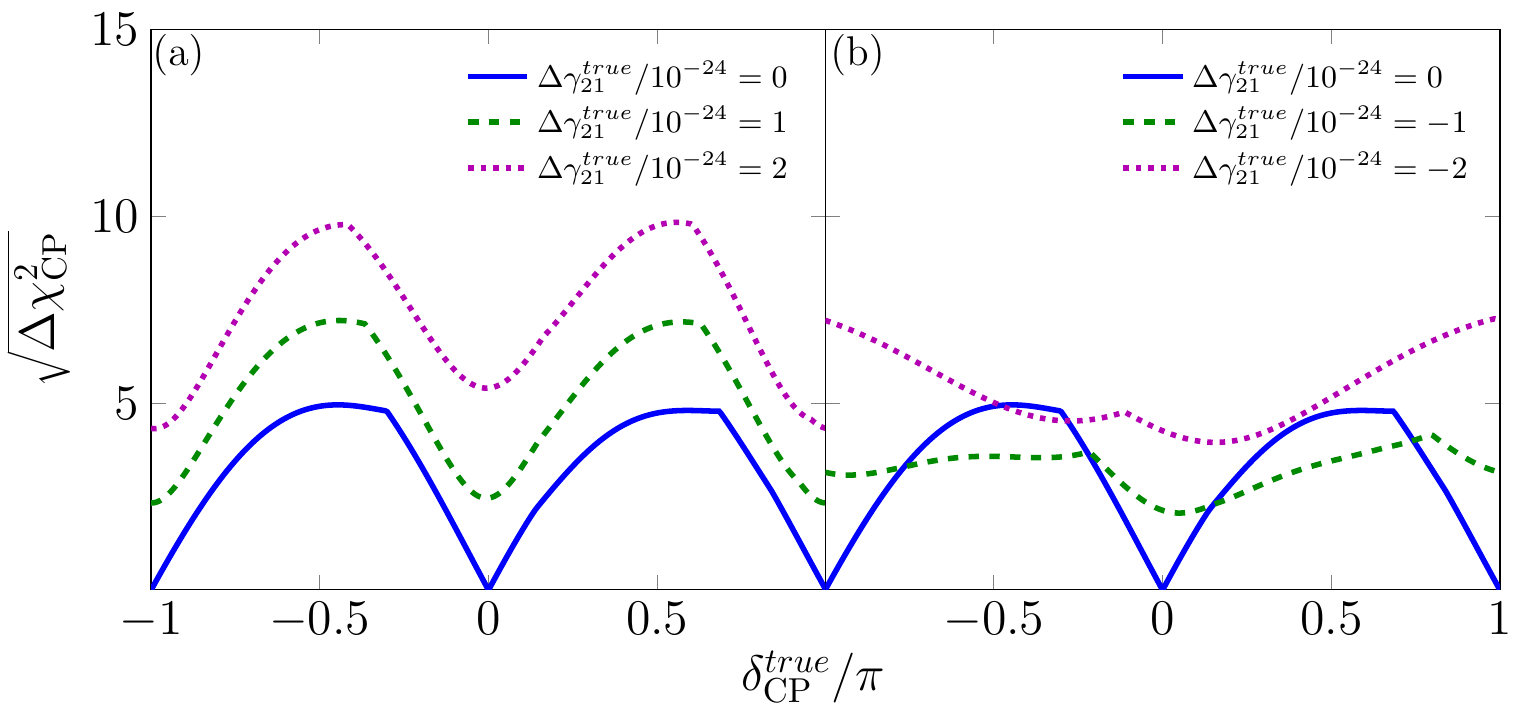}
	\caption{CP Violation sensitivity for scenario A/case 2.}
	\label{CPV_sensi_scA_cs2_panel}
\end{figure}

In Fig. \ref{CPV_sensi_scA_cs2_panel} (a), scenario A/case 2, the displayed results are due to the increased asymmetry between the $\nu_\mu \rightarrow \nu_e$ and $\bar{\nu}_\mu \rightarrow \bar{\nu}_e$ appearance channels amplifying the discrimination of the CP violation case, see Fig. \ref{probsAc2mm}. This also includes an extra fake CPV caused by the connection between the VEP term and the matter potential. Notwithstanding, as a consequence of the opposite behavior (decrease) of the asymmetry  between the $\nu_\mu \rightarrow \nu_e$ and $\bar{\nu}_\mu \rightarrow \bar{\nu}_e$ appearance channels, when $\Delta \gamma_{21}<0$, it is observed a decrease in the level of significance, that could be even lower to the SO case in the neighborhood of $\delta_{\mathrm{CP}}^{true}=\pm \pi/2$, where this
case reaches its peak of sensitivity. This means that the capacity to reject the null CP-hypothesis when $\delta_{\mathrm{CP}}^{true}$ takes values close to its maximum would be reduced. As already stated,  the results for scenario A/case 2 are qualitatively similar to those shown in scenario B/texture $\theta_{12}$. Moreover, scenario B/texture $\theta_{13}$  $\Delta \gamma_{31}>0 \ (\Delta \gamma_{31}<0)$ corresponds to $\Delta \gamma_{21}<0 \ (\Delta \gamma_{21}>0)$ for scenario A/case 2. Therefore, we could apply Fig.~\ref{CPV_sensi_scA_cs2_panel} and explanations for scenario A/case 2 to these ones.

\subsubsection{Mass Hierarchy Sensitivity}

One of the main goals of DUNE experiment is to figure out the mass hierarchy (MH). This is related to the fact that one of the main features of DUNE experiment is its baseline (1300 Km),  resulting in a high sensitivity to the matter effect. This means that a considerable difference in the oscillation channels $\nu_\mu \rightarrow \nu_e$ and $\bar{\nu}_\mu \rightarrow \bar{\nu}_e$ is expected as a result, on which MH depends. Therefore, studying the sensitivity to MH is extremely important, since we have shown VEP scenarios where the asymmetry of these channels is clearly affected. The MH sensitivity is obtained as follows \cite{Acciarri:2015uup,Carpio:2018gum}.

\begin{equation}
\begin{split}
\Delta \chi_{\mathrm{MH}}^2 =&  \chi^2 ({\Delta m_{31}^2}^{test}<0,\Delta \gamma^{test}=0, \\ 
&{\Delta m_{31}^2}^{true}>0,\delta_{\mathrm{CP}}^{true},\Delta \gamma^{true}) \\
\end{split}
\end{equation}
Taking into account the analysis explained in the previous section we study the impact on the MH sensitivity 
considering VEP/NH in nature and assuming SO/IH as theoretical hypothesis. We do not display the scenarios with low discrepancies on 
 $\nu_\mu \rightarrow \nu_e$, which are scenario A/case 1 and scenario B/texture $\theta_{23}$-a and texture $\theta_{23}$-b since those scenarios have MH sensitivities rather similar to SO MH.

\begin{figure}[h!]
	\includegraphics[scale=0.55]{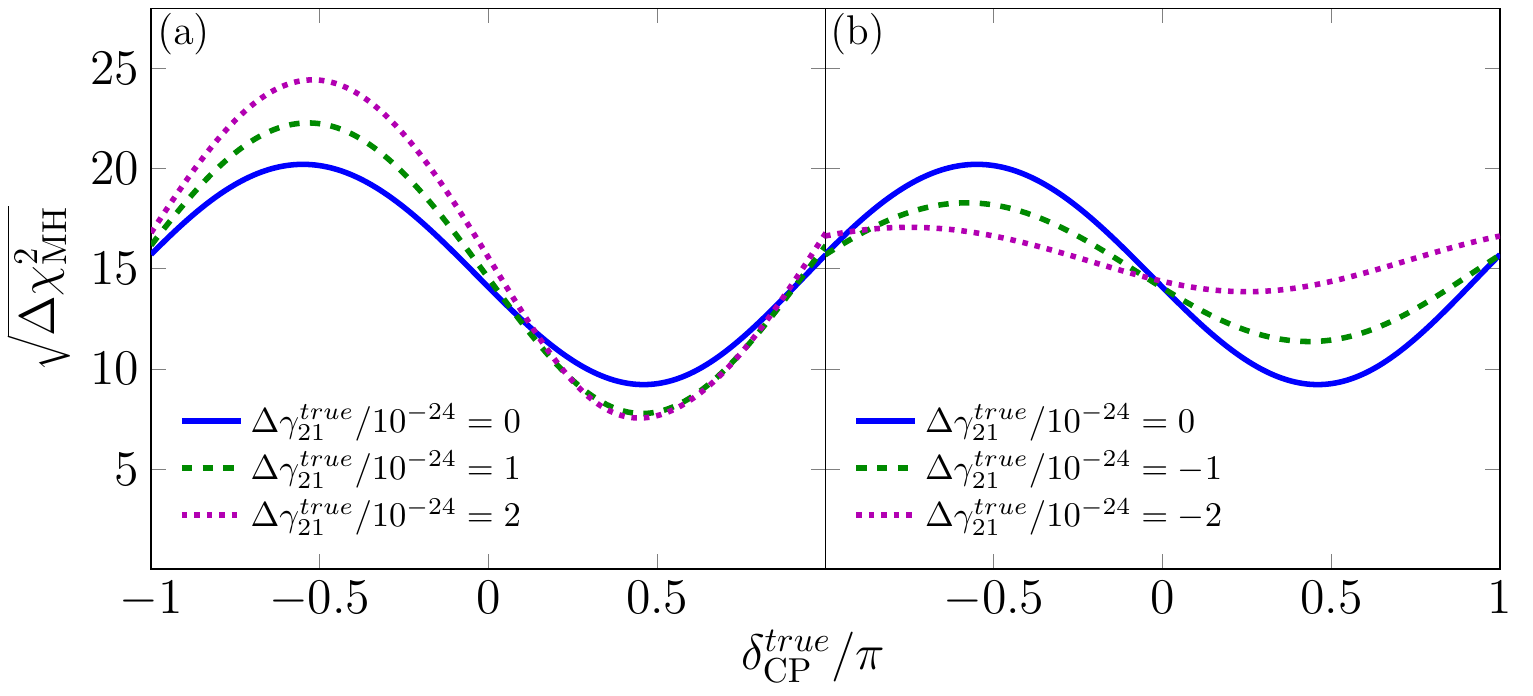}
	\caption{Mass Hierarchy sensitivity for scenario A/case 2.}
	\label{MH_sensi_scA_cs2_panel}
\end{figure}

In Fig. \ref{MH_sensi_scA_cs2_panel} the MH sensitivities for scenario A/case 2 are presented.  In order to explain the behavior of these sensitivity curves we define two probability differences:  
$\Delta P^{\text{SO}}= P_{\nu_\mu \rightarrow \nu_e}^{\text{SO(NH)}} -P_{\nu_\mu \rightarrow \nu_e }^{\text{SO(IH)}}$ and $\Delta P^{\text{VEP}}= P_{\nu_\mu \rightarrow \nu_e}^{\text{VEP}\bigoplus \text{SO(NH)}} -P_{\nu_\mu \rightarrow \nu_e }^{\text{SO(IH)}}$ with
$\Delta^{\text{VEP-SO(NH)}}=\Delta P^{\text{VEP}} - \Delta P^{\text{SO}}$. The $\Delta P^{\text{VEP}}$ is associated with the VEP sensitivity while $\Delta P^{\text{SO}}$ is related to the SO one. For this scenario the most important VEP-terms of $s_{13} \Delta \tilde{\gamma}_{21}\sim  \mathcal{O}(0.01)$ of the transition probability (see Eq.~(\ref{eqs1c2me})) can be written into a single term proportional to $ fg \Delta \tilde{\gamma}_{21}$, considering $\Delta \sim \pi/2$. For $\Delta \gamma_{21}>0$, $\Delta^{\text{VEP-SO(NH)}} \propto fg \Delta \tilde{\gamma}_{21}$ at $\delta_{\mathrm{CP}}^{true} = -\pi/2$, therefore the VEP sensitivity reaches a higher significance than the SO one. While, at $\delta_{\mathrm{CP}}^{true} = \pi/2$, $\Delta^{\text{VEP-SO(NH)}}\propto -fg \Delta \tilde{\gamma}_{21}$, which means that the VEP sensitivity attains lower significance than the SO one. For $\Delta \gamma_{21}<0$, what happens is exactly the opposite. These results are applicable for scenario B/texture $\theta_{13}$  and texture $\theta_{12}$, as well.

\section{Conclusions}
We have tested the impact of fitting simulated data generated for different VEP scenarios, and considering pure standard oscillation as theoretical hypothesis. Among our findings, we have found the displacement of the $\Delta m^2_{31}$,  the increase of $\sin^2 \theta_{13}$ ($\Delta \gamma >0$) or the change of $\delta_{\text{CP}}$ ($\Delta \gamma >0$) toward the decrease of the magnitude of CP violation, which are scenario-dependent effects. Furthermore, the DUNE CP sensitivity, treating VEP as before, increases for the majority of scenarios having all in common the introduction of a fake CP violation. The DUNE significance for identifying the MH for $\Delta \gamma >0$  ($\Delta \gamma< 0$) increases (decreases) and decreases (increases) for $\delta_{\text{CP}} \in [-\pi,0]$ and $\delta_{\text{CP}} \in [0,\pi]$. In addition, we have also found limits for VEP, for the variety of scenarios under study, being the most stringent $\Delta \gamma \sim  0.7 \times 10^{-24}$ GeV which corresponds to the scenario B/texture $\theta_{12}$. Finally, we have set limits for LV terms of the SME Hamiltonian, with different energy dependencies, the most restrictive one corresponds to the scenario B/texture $\theta_{12}$, as well, and is $\Delta \gamma = \{ 8, 1.5, 0.12, 0.007 \} \times 10^{-24}$ GeV that corresponds to $n=0,1,2,3$, respectively.

\section{Acknowledgements}

 A. M. Gago acknowledges funding by the {\it Direcci\'on de Gesti\'on de la Investigaci\'on} at PUCP, through grants  DGI-2017-3-0019 and DGI 2019-3-0044. F. N. D\'iaz acknowledges CONCYTEC for the graduate  fellowship  under  Grant  No. 000236-2015-FONDECYT-DE. The authors also want to thank F. de Zela and J. L. Bazo for useful suggestions and reading the manuscript.


\begin{thebibliography}{0}
\expandafter\ifx\csname natexlab\endcsname\relax\def\natexlab#1{#1}\fi
\expandafter\ifx\csname bibnamefont\endcsname\relax
  \def\bibnamefont#1{#1}\fi
\expandafter\ifx\csname bibfnamefont\endcsname\relax
  \def\bibfnamefont#1{#1}\fi
\expandafter\ifx\csname citenamefont\endcsname\relax
  \def\citenamefont#1{#1}\fi
\expandafter\ifx\csname url\endcsname\relax
  \def\url#1{\texttt{#1}}\fi
\expandafter\ifx\csname urlprefix\endcsname\relax\def\urlprefix{URL }\fi
\providecommand{\bibinfo}[2]{#2}
\providecommand{\eprint}[2][]{\url{#2}}

\end{thebibliography}


\begin{thebibliography}{}


\bibitem{Fukuda:1998mi} 
  Y.~Fukuda {\it et al.} [Super-Kamiokande Collaboration],
  \href{https://doi.org/10.1103/PhysRevLett.81.1562}{Phys.\ Rev.\ Lett.\  {\bf 81}, 1562 (1998)}


\bibitem{Fukuda:2001nj} 
  S.~Fukuda {\it et al.} [Super-Kamiokande Collaboration],
  \href{https://doi.org/10.1103/PhysRevLett.86.5651}{Phys.\ Rev.\ Lett.\  {\bf 86}, 5651 (2001)}


\bibitem{Ahmad:2002jz} 
  Q.~R.~Ahmad {\it et al.} [SNO Collaboration],
  \href{https://doi.org/10.1103/PhysRevLett.89.011301}{Phys.\ Rev.\ Lett.\  {\bf 89}, 011301 (2002)}



\bibitem{Araki:2004mb} 
  T.~Araki {\it et al.} [KamLAND Collaboration],
  \href{https://doi.org/10.1103/PhysRevLett.94.081801}{Phys.\ Rev.\ Lett.\  {\bf 94}, 081801 (2005)}


\bibitem{Adamson:2007gu} 
  P.~Adamson {\it et al.} [MINOS Collaboration],
  \href{https://doi.org/10.1103/PhysRevD.77.072002}{Phys.\ Rev.\ D {\bf 77}, 072002 (2008)}



\bibitem{An:2012eh} 
  F.~P.~An {\it et al.} [Daya Bay Collaboration],
  \href{https://doi.org/10.1103/PhysRevLett.108.171803}{Phys.\ Rev.\ Lett.\  {\bf 108}, 171803 (2012)}


\bibitem{Ahn:2012nd} 
  J.~K.~Ahn {\it et al.} [RENO Collaboration],
  \href{https://doi.org/10.1103/PhysRevLett.108.191802}{Phys.\ Rev.\ Lett.\  {\bf 108}, 191802 (2012)}



\bibitem{Abe:2011fz} 
  Y.~Abe {\it et al.} [Double Chooz Collaboration],
  \href{https://doi.org/doi:10.1103/PhysRevLett.108.131801}{Phys.\ Rev.\ Lett.\  {\bf 108}, 131801 (2012)}


\bibitem{Kajita:2016vhj} 
  T.~Kajita {\it et al.} [Super-Kamiokande Collaboration],
  \href{https://doi.org/10.1016/j.nuclphysb.2016.04.017}{Nucl.\ Phys.\ B {\bf 908}, 14 (2016).)}





\bibitem{Gago:2001xg} 
  A.~M.~Gago, M.~M.~Guzzo, H.~Nunokawa, W.~J.~C.~Teves and R.~Zukanovich Funchal,
 \href{https://doi.org/10.1103/PhysRevD.64.073003}{Phys.\ Rev.\ D {\bf 64}, 073003 (2001)}

\bibitem{Guzzo:2004ue} 
  M.~M.~Guzzo, P.~C.~de Holanda and O.~L.~G.~Peres,
  \href{https://doi.org/10.1016/j.physletb.2004.04.035}{Phys.\ Lett.\ B {\bf 591}, 1 (2004)}
  

  
  \bibitem{deGouvea:2015ndi} 
  A.~de Gouvêa and K.~J.~Kelly,
  \href{https://doi.org/10.1016/j.nuclphysb.2016.03.013}{Nucl.\ Phys.\ B {\bf 908}, 318 (2016)}

\bibitem{Masud:2016gcl} 
  M.~Masud and P.~Mehta,
  \href{https://doi.org/10.1103/PhysRevD.94.053007}{Phys.\ Rev.\ D {\bf 94}, no. 5, 053007 (2016)}

\bibitem{Liao:2016orc} 
  J.~Liao, D.~Marfatia and K.~Whisnant,
  \href{https://doi.org/10.1007/JHEP01(2017)071}{JHEP {\bf 1701}, 071 (2017)}




\bibitem{Frieman:1987as} 
  J.~A.~Frieman, H.~E.~Haber and K.~Freese,
  \href{https://doi.org/10.1016/0370-2693(88)91120-3}{Phys.\ Lett.\ B {\bf 200}, 115 (1988).}

\bibitem{Barger:1999bg} 
  V.~D.~Barger, J.~G.~Learned, P.~Lipari, M.~Lusignoli, S.~Pakvasa and T.~J.~Weiler,
  \href{https://doi.org/10.1016/S0370-2693(99)00887-4}{Phys.\ Lett.\ B {\bf 462}, 109 (1999)}

\bibitem{Bandyopadhyay:2002qg} 
  A.~Bandyopadhyay, S.~Choubey and S.~Goswami,
  \href{https://doi.org/10.1016/S0370-2693(03)00044-3}{Phys.\ Lett.\ B {\bf 555}, 33 (2003)}

\bibitem{Fogli:2004gy} 
  G.~L.~Fogli, E.~Lisi, A.~Mirizzi and D.~Montanino,
  \href{https://doi.org/10.1103/PhysRevD.70.013001}{Phys.\ Rev.\ D {\bf 70}, 013001 (2004)}



\bibitem{Berryman:2014yoa} 
  J.~M.~Berryman, A.~de Gouv\^ea, D.~Hernández and R.~L.~N.~Oliveira,
  \href{https://doi.org/10.1016/j.physletb.2015.01.002}{Phys.\ Lett.\ B {\bf 742}, 74 (2015)}


\bibitem{Picoreti:2015ika} 
  R.~Picoreti, M.~M.~Guzzo, P.~C.~de Holanda and O.~L.~G.~Peres,
  \href{https://doi.org/10.1016/j.physletb.2016.08.007}{Phys.\ Lett.\ B {\bf 761}, 70 (2016)}


\bibitem{Bustamante:2016ciw} 
  M.~Bustamante, J.~F.~Beacom and K.~Murase,
  \href{https://doi.org/10.1103/PhysRevD.95.063013}{Phys.\ Rev.\ D {\bf 95}, no. 6, 063013 (2017)}

\bibitem{Gago:2017zzy} 
  A.~M.~Gago, R.~A.~Gomes, A.~L.~G.~Gomes, J.~Jones-Perez and O.~L.~G.~Peres,
  \href{https://doi.org/10.1007/JHEP11(2017)022}{JHEP {\bf 1711}, 022 (2017)}
 
\bibitem{Ascencio-Sosa:2018lbk} 
  M.~V.~Ascencio-Sosa, A.~M.~Calatayud-Cadenillas, A.~M.~Gago and J.~Jones-P\'erez,
  \href{https://doi.org/10.1140/epjc/s10052-018-6276-0}{Eur.\ Phys.\ J.\ C {\bf 78}, no. 10, 809 (2018)}
  
\bibitem{deSalas:2018kri} 
  P.~F.~de Salas, S.~Pastor, C.~A.~Ternes, T.~Thakore and M.~Tórtola,
  \href{https://doi.org/10.1016/j.physletb.2018.12.066}{Phys.\ Lett.\ B {\bf 789}, 472 (2019)}







\bibitem{Lisi:2000zt} 
  E.~Lisi, A.~Marrone and D.~Montanino,
  \href{https://doi.org/10.1103/PhysRevLett.85.1166}{Phys.\ Rev.\ Lett.\  {\bf 85}, 1166 (2000)}

\bibitem{Barenboim:2006xt} 
  G.~Barenboim, N.~E.~Mavromatos, S.~Sarkar and A.~Waldron-Lauda,
  \href{https://doi.org/10.1016/j.nuclphysb.2006.09.012}{Nucl.\ Phys.\ B {\bf 758}, 90 (2006)}

\bibitem{Bakhti:2015dca} 
  P.~Bakhti, Y.~Farzan and T.~Schwetz,
  \href{https://doi.org/10.1007/JHEP05(2015)007}{JHEP {\bf 1505}, 007 (2015)}


\bibitem{Carpio:2017nui} 
  J.~A.~Carpio, E.~Massoni and A.~M.~Gago,
  \href{https://doi.org/10.1103/PhysRevD.97.115017}{Phys.\ Rev.\ D {\bf 97}, no. 11, 115017 (2018)}


\bibitem{Capolupo:2018hrp} 
  A.~Capolupo, S.~M.~Giampaolo and G.~Lambiase,
  \href{https://doi.org/10.1016/j.physletb.2019.03.062}{Phys.\ Lett.\ B {\bf 792}, 298 (2019)}

\bibitem{Carrasco:2018sca} 
  J.~C.~Carrasco, F.~N.~D\'iaz and A.~M.~Gago,
  \href{https://doi.org/10.1103/PhysRevD.99.075022}{Phys.\ Rev.\ D {\bf 99}, no. 7, 075022 (2019)}

\bibitem{Gomes:2020muc} 
  A.~L.~G.~Gomes, R.~A.~Gomes and O.~L.~G.~Peres,
  \href{https://arxiv.org/pdf/2001.09250.pdf}{arXiv:2001.09250 [hep-ph].}



  
\bibitem{Adamson:2008aa} 
  P.~Adamson {\it et al.} [MINOS Collaboration],
  \href{https://doi.org/10.1103/PhysRevLett.101.151601}{Phys.\ Rev.\ Lett.\  {\bf 101}, 151601 (2008)}
  
\bibitem{AguilarArevalo:2011yi} 
  A.~A.~Aguilar-Arevalo {\it et al.} [MiniBooNE Collaboration],
  \href{https://doi.org/10.1016/j.physletb.2012.12.020}{Phys.\ Lett.\ B {\bf 718}, 1303 (2013)}
  
  
\bibitem{Li:2014rya} 
  Y.~F.~Li and Z.~h.~Zhao,
  \href{https://doi.org/10.1103/PhysRevD.90.113014}{Phys.\ Rev.\ D {\bf 90}, no. 11, 113014 (2014)}







\bibitem{Gasperini:1988zf} 
  M.~Gasperini,
  \href{https://doi.org/10.1103/PhysRevD.38.2635}{Phys.\ Rev.\ D {\bf 38}, 2635 (1988).}



\bibitem{Gasperini:1989rt} 
  M.~Gasperini,
  \href{https://doi.org/10.1103/PhysRevD.39.3606}{Phys.\ Rev.\ D {\bf 39}, 3606 (1989).}


	
	
	\bibitem{Halprin:1991gs} 
  A.~Halprin and C.~N.~Leung,
  \href{https://doi.org/10.1103/PhysRevLett.67.1833}{Phys.\ Rev.\ Lett.\  {\bf 67}, 1833 (1991).}
  
  
  \bibitem{Pantaleone:1992ha} 
  J.~T.~Pantaleone, A.~Halprin and C.~N.~Leung,
  \href{https://doi.org/10.1103/PhysRevD.47.R4199}{Phys.\ Rev.\ D {\bf 47}, R4199 (1993)}
  
  \bibitem{Butler:1993wi} 
  M.~N.~Butler, S.~Nozawa, R.~A.~Malaney and A.~I.~Boothroyd,
  \href{https://doi.org/10.1103/PhysRevD.47.2615}{Phys.\ Rev.\ D {\bf 47}, 2615 (1993).}
  
  \bibitem{Bahcall:1994zw} 
  J.~N.~Bahcall, P.~I.~Krastev and C.~N.~Leung,
  \href{https://doi.org/10.1103/PhysRevD.52.1770}{Phys.\ Rev.\ D {\bf 52}, 1770 (1995)}
  
  
 




\bibitem{Mansour:1998nb} 
  S.~W.~Mansour and T.~K.~Kuo,
  \href{https://doi.org/10.1103/PhysRevD.60.097301}{Phys.\ Rev.\ D {\bf 60}, 097301 (1999)}
  
  
  \bibitem{Gago:1999hi} 
  A.~M.~Gago, H.~Nunokawa and R.~Zukanovich Funchal,
  \href{https://doi.org/10.1103/PhysRevLett.84.4035}{Phys.\ Rev.\ Lett.\  {\bf 84}, 4035 (2000)}




\bibitem{Yasuda:1994nu} 
  O.~Yasuda,
  \href{https://arxiv.org/pdf/gr-qc/9403023.pdf}{gr-qc/9403023.}


\bibitem{Datta:2000hm} 
  A.~Datta,
  \href{https://doi.org/10.1016/S0370-2693(01)00299-4}{Phys.\ Lett.\ B {\bf 504}, 247 (2001)}




	


\bibitem{valdiviessotesis}
	G. D. A. Valdiviesso, \href{http://repositorio.unicamp.br/jspui/bitstream/REPOSIP/278255/1/Valdiviesso_GustavodoAmaral_D.pdf}{{\it Novos limites para violação do princ\'ipio da equival\^encia em neutrinos solares}}, Ph. D. thesis in portuguese, University of Campinas, Brazil, 2008.





	



\bibitem{Valdiviesso:2012nva} 
  G.~A.~Valdiviesso, M.~M.~Guzzo and P.~C.~Holanda,
  \href{https://doi.org/10.1016/j.nuclphysbps.2012.09.089}{Nucl.\ Phys.\ Proc.\ Suppl.\  {\bf 229-232}, 452 (2012).}

\bibitem{Esmaili:2014ota} 
  A.~Esmaili, D.~R.~Gratieri, M.~M.~Guzzo, P.~C.~de Holanda, O.~L.~G.~Peres and G.~A.~Valdiviesso,
  \href{https://doi.org/10.1103/PhysRevD.89.113003}{Phys.\ Rev.\ D {\bf 89}, no. 11, 113003 (2014)}
	

	
	
	





\bibitem{Alion:2016uaj} 
  T.~Alion {\it et al.} [DUNE Collaboration],
  \href{https://arxiv.org/pdf/1606.09550.pdf}{arXiv:1606.09550 [physics.ins-det]}.

	
\bibitem{Acciarri:2015uup} 
  R.~Acciarri {\it et al.} [DUNE Collaboration],
  \href{https://arxiv.org/pdf/1512.06148.pdf}{arXiv:1512.06148 [physics.ins-det].}
  
  	






\bibitem{Colladay:1996iz} 
  D.~Colladay and V.~A.~Kostelecky,
  \href{https://doi.org/10.1103/PhysRevD.55.6760}{Phys.\ Rev.\ D {\bf 55}, 6760 (1997)}


\bibitem{Colladay:1998fq} 
  D.~Colladay and V.~A.~Kostelecky,
  \href{https://doi.org/10.1103/PhysRevD.58.116002}{Phys.\ Rev.\ D {\bf 58}, 116002 (1998)}

\bibitem{Kostelecky:1988zi} 
  V.~A.~Kostelecky and S.~Samuel,
  \href{https://doi.org/10.1103/PhysRevD.39.683}{Phys.\ Rev.\ D {\bf 39}, 683 (1989).}
		
		
\bibitem{Liao:2016hsa} 
  J.~Liao, D.~Marfatia and K.~Whisnant,
  \href{https://doi.org/10.1103/PhysRevD.93.093016}{Phys.\ Rev.\ D {\bf 93}, no. 9, 093016 (2016)}


\bibitem{Majhi:2019tfi} 
  R.~Majhi, C.~Soumya and R.~Mohanta,
  \href{https://arxiv.org/pdf/1907.09145.pdf}{arXiv:1907.09145 [hep-ph].}
	
\bibitem{Nufit}
	\href{http://www.nu-fit.org}{http://www.nu-fit.org}

	
	
\bibitem{Aartsen:2017ibm} 
  M.~G.~Aartsen {\it et al.} [IceCube Collaboration],
  \href{https://doi.org/10.1038/s41567-018-0172-2}{Nature Phys.\  {\bf 14}, no. 9, 961 (2018)}
	
	
	
	
	\bibitem{Huber:2004ka} 
  P.~Huber, M.~Lindner and W.~Winter,
  \href{https://doi.org/10.1016/j.cpc.2005.01.003}{Comput.\ Phys.\ Commun.\  {\bf 167}, 195 (2005)}
	
	
\bibitem{Huber:2007ji} 
  P.~Huber, J.~Kopp, M.~Lindner, M.~Rolinec and W.~Winter,
  \href{https://doi.org/10.1016/j.cpc.2007.05.004}{Comput.\ Phys.\ Commun.\  {\bf 177}, 432 (2007)}
	
\bibitem{Carpio:2018gum} 
  J.~A.~Carpio, E.~Massoni and A.~M.~Gago,
  \href{https://doi.org/10.1103/PhysRevD.100.015035}{Phys.\ Rev.\ D {\bf 100}, no. 1, 015035 (2019)}
	

	
\bibitem{Jurkovich:2018rif} 
  H.~Jurkovich, C.~P.~Ferreira and P.~Pasquini,
  \href{https://arxiv.org/pdf/1806.08752.pdf}{arXiv:1806.08752 [hep-ph].}
	
	
	\bibitem{Barenboim:2018ctx} 
  G.~Barenboim, M.~Masud, C.~A.~Ternes and M.~Tórtola,
  \href{https://doi.org/10.1016/j.physletb.2018.11.040}{Phys.\ Lett.\ B {\bf 788}, 308 (2019)}


	


	
\end{thebibliography}
\end{document}